\documentclass[aps,pra,twocolumn,groupedaddress,amsmath,amssymb,superscriptaddress]{revtex4-2}
\usepackage{graphicx}  % needed for figures
\usepackage{dcolumn}   % needed for some tables
\usepackage{bm}        % for math
\usepackage{verbatim}   % for math
\usepackage{ mathrsfs }
\usepackage[colorlinks=true,linkcolor=blue,citecolor=blue,allcolors=blue]{hyperref}
\numberwithin{equation}{section}
\usepackage{float}

\begin{document}

\title{Molecular excited state in the interaction quench dynamics of two different atoms in a two-dimensional anisotropic trap}

\author{I. S. Ishmukhamedov}
\email{i.ishmukhamedov@mail.ru}
\affiliation{Institute of Nuclear Physics, Almaty 050032, Kazakhstan}
\affiliation{Al-Farabi Kazakh National University, Almaty 050040, Kazakhstan}

\author{A. S. Ishmukhamedov}
\affiliation{Institute of Nuclear Physics, Almaty 050032, Kazakhstan}
\affiliation{Al-Farabi Kazakh National University, Almaty 050040, Kazakhstan}

\author{Zh. E. Jalankuzov}
\affiliation{Institute of Nuclear Physics, Almaty 050032, Kazakhstan}
\affiliation{Al-Farabi Kazakh National University, Almaty 050040, Kazakhstan}

\author{D. V. Ismailov}
\affiliation{Al-Farabi Kazakh National University, Almaty 050040, Kazakhstan}
\affiliation{K. I. Satpayev Kazakh National Technical University, Almaty 050013, Kazakhstan}
%\date{\today}
\begin{abstract}
We explore the interaction quench dynamics of two atoms with different masses and subject to different trapping potentials. Notably, under such anisotropic conditions, the nonequilibrium dynamics can lead to the occupation of molecular excited states. We consider cases of quenching from attractive to repulsive interaction and vice versa, analyzing the impact of the pre- and postquench states. The analysis of overlap integrals for the both states reveals a significant contribution from the molecular excited state. Moreover, the overlap with the prequench states might serve as an indicator of when this excited state may emerge. Additionally, we calculate the energy spectrum for the lowest levels in the both isotropic and anisotropic harmonic traps. Throughout our study, we use a Gaussian-shaped finite-range interaction potential.
\end{abstract}

\maketitle

\section{Introduction}

Ultracold atomic physics represents an intriguing and highly promising field of research. The capacity to manipulate particle interactions, adjust system dimensionality, and prepare well-defined quantum states enables the modeling of diverse complex systems with great significance. Among the intriguing applications of such systems, quantum computing stands out. Various approaches \cite{hartke,wu,bruzewicz} have already been explored, utilizing two-dimensional geometries to create fundamental building blocks for quantum computers.

A prevalent approximation underlying numerous approaches is the assumption of a harmonic oscillator potential for the confining trap. This approximation has paved the way for many analytical models that play a crucial role in describing ultracold systems across various dimensions and particle numbers. The work of Busch et al. \cite{busch} is particularly noteworthy, as they successfully derived exact solutions for two atoms trapped in an isotropic harmonic oscillator potential across 1D, 2D, and 3D geometries. Additionally, there are other works that provide exact analytical solutions for axially symmetric traps of different geometries \cite{idziaszek}, as well as completely anisotropic traps \cite{chen}.

In the case of narrow transverse confinement, Olshanii \cite{olshanii} discovered the phenomenon known as confinement-induced resonance (CIR). CIR occurs when the energy of two colliding atoms reaches the threshold of a transversely excited molecular bound state. Further investigations \cite{melezhik2007,melezhik2009} revealed that the coupling between center-of-mass and relative motions plays a crucial role in the mechanism of molecule formation. This was confirmed experimentally in \cite{haller}, where CIR in anisotropic transversal confinement was investigated. Interestingly, additional resonances were observed in their experiments that could not be explained by the conventional CIR model at that time. Subsequently, the nature of these new CIRs was attributed to the coupling between center-of-mass and relative motions \cite{sala,peng,sala2}. This coupling can arise due to trap anisotropy or anharmonicity. The study of tunneling and nonequilibrium dynamics in anharmonic traps has uncovered interesting physics, which were explored in \cite{gharashi, dobrzyniecki, ishmukh, ishmukh2, ishmukh3}.

In a remarkable experiment \cite{serwane}, a setup was developed capable of trapping only a few atoms, including just two. This opened up a way for experimental realization and validation of quantum theories on a microscopic scale to an unprecedented level of reliability. One of the primary focuses of the early investigations \cite{zuern} was the examination of the fermionization effect. This effect is observed when the energy and square modulus of the wave function for two impenetrable bosons are identical to those of two noninteracting identical fermions.  In a two-dimensional geometry, researchers were able to directly investigate the quantum phase transition from a normal state to a superfluid state, examining this transition from the perspective of few-body systems \cite{bayha,holten}.

Nonequilibrium dynamics governs diverse fields of research ranging from quantum information up to cosmology \cite{langen}. With the advent of ultracold atoms, such problems can be tackled in a new, more efficient way.

Thus, addressing the quench dynamics of two interacting atoms in a two-dimensional geometry becomes a natural step towards exploring any of the above mentioned directions. Following \cite{melezhik2007,melezhik2009} we consider two atoms with different masses and confinement potentials. We study the interaction quench dynamics in a two-dimensional geometry as in \cite{bougas} (which we will frequently cite throughout the paper), except here we use an anisotropic trap and a finite-range interaction potential of a Gaussian shape. To ensure consistency, we adjust the strength of the interaction potential to reproduce the same energy levels as those obtained with the zero-range interaction potential, so that we can work in terms of the coupling strength of the contact interaction. By comparing our computed fidelity with the fidelity computed using the zero-range interaction potential, we find a close agreement. A similar adjustment between the two types of interaction potentials in a one-dimensional geometry was explored in \cite{gharashi, ishmukh4}. Additionally, reference \cite{ishmukh5} employed a numerical approximation of the zero-range interaction to study the energy spectrum.

In the case of the anisotropic trap, the analysis of overlap integrals with the pre- and post-quench states, as well as the evolution of the probability density, reveals that the system occasionally occupies a molecular excited state during the dynamics.

The remainder of this paper is organized as follows. Section \ref{model} introduces the model Hamiltonian and the method employed for the solution of the corresponding Schr\"{o}dinger equation. In Section \ref{spectrum}, we present the spectrum for the lowest energy levels in both isotropic and anisotropic harmonic traps. Section \ref{isotropic} focuses on comparing our results for the quench dynamics in the isotropic trap, considering both the finite-range and zero-range interaction potentials. Section  \ref{anisotropic} explores the quench dynamics in the anisotropic trap. Finally, we summarize our findings in the concluding Section \ref{summary}.

\section{Model Hamiltonian and the method}\label{model}

We consider the Hamiltonian of two atoms with masses $m_1$ and $m_2$ on a plane:
\begin{equation}\label{ham}
H=-\frac{\hbar^2}{2m_1}\Delta_1-\frac{\hbar^2}{2m_2}\Delta_2
+V(\rho_1)+V(\rho_2)+V_{\textmd{int}}(\bm{\rho}_1-\bm{\rho}_2),
\end{equation}
where the Laplacian in cylindrical coordinates reads:
\begin{equation}\label{laplacian}
\Delta_j=\frac{\partial^2}{\partial \rho_j^2}+\frac{1}{\rho_j}\frac{\partial}{\partial \rho_j}-\frac{1}{\rho_j^2}L_{z_j}^2,~~~~j=1,2,
\end{equation}
where $L_{z_j}=-i\hbar \partial/\partial \varphi_j$
The confining trap for the both particles is given by:
\begin{equation}\label{trap}
V(\rho_j)=\frac{1}{2}m_j\omega_j^2\rho_j^2,~~~~j=1,2
\end{equation}
We choose the Gaussian interaction potential:
\begin{equation}\label{intgs}
V_{\textmd{int}}(\bm{\rho}_1-\bm{\rho}_2)=V_0\exp\left\{ -\dfrac{(\bm{\rho}_1-\bm{\rho}_2)^2 }{2\rho_0^2} \right\},
\end{equation}
The factor of 2 in the exponent is used for convenience. 

From here on, we will use oscillator units, where energy is given in $\hbar\omega_0$, length is measured in units of $\sqrt{\hbar/(m_0\omega_0)}$, time is measured in units of $\omega_0^{-1}$, and mass is measured in units of $m_0$. The width of the Gaussian interaction potential \eqref{intgs} in the oscillator units is set to $\rho_0=0.1$, which is deemed sufficient to adequately model the interaction between atoms \cite{doganov}.

Now, we will make a transition to the relative coordinates $\bm{r}(r,\varphi)$ and the center-of-mass coordinates $\bm{R}(R,\phi)$ as follows:
\begin{equation}
	\begin{cases}
		\bm{r}=\dfrac{\bm{\rho}_1-\bm{\rho}_2}{\sqrt{2}} \\
		\bm{R}=\sqrt{2}\dfrac{m_1\bm{\rho}_1+m_2\bm{\rho}_2}{m_1+m_2}
	\end{cases}.
\end{equation}
We found that introducing the factor $\sqrt{2}$ provides slightly better computational stability.

In the new variables, the Hamiltonian is given by:
\begin{equation}\label{ham2}
	H=-\frac{\hbar^2}{4\mu}\Delta_r-\frac{\hbar^2}{\mathcal{M}}\Delta_R
	+V_1(r)+V_2(R)+W(r,R)+V_{\textmd{int}}(r),
\end{equation}
where $\mu=m_1m_2/(m_1+m_2)$ and $\mathcal{M}=m_1+m_2$, and
\begin{equation}
	V_1(r)=\mu^2\left(\frac{\omega_1^2}{m_1}+\frac{\omega_2^2}{m_2}\right)r^2,
\end{equation}
\begin{equation}
	V_2(R)=\frac{1}{4}\left(m_1\omega_1^2+m_2\omega_2^2\right)R^2,
\end{equation}
and
\begin{equation}
	W(r,R)=\mu\left(\omega_1^2-\omega_2^2\right)rR\cos\varphi
\end{equation}
Next, we employ the result from \cite{melezhik2007,melezhik2009,bock}, where it is possible to eliminate the angular part of the center-of-mass motion by performing a unitary rotating-frame transformation of the Hamiltonian \eqref{ham2}. Due to the conservation of the total angular momentum $L_z+L_Z$, it is possible to remove the angular dependence on $\phi$. The resulting Hamiltonian reads:
\begin{equation}\label{ham3}
	\begin{aligned}
	H&=-\frac{1}{4\mu}\left(\frac{\partial^2}{\partial r^2}+\frac{1}{r}\frac{\partial}{\partial r}\right)
	-\frac{1}{\mathcal{M}}\left(\frac{\partial^2}{\partial R^2}+\frac{1}{R}\frac{\partial}{\partial R}\right) \\
	&-\left(\frac{1}{4\mu r^2}+\frac{1}{\mathcal{M} R^2}\right)\frac{\partial^2}{\partial \varphi^2}
	+V_1(r)+V_2(R)\\
	&+W(r,R)+V_{\textmd{int}}(r),
	\end{aligned}
\end{equation}
we consider only the case of the zero value $M_{\phi}=0$ for the quantum number of the operator $L_Z$.
Now, we proceed to transform the wave function as follows \cite{borisov,lemoine}:
\begin{equation}
	\Psi(r,R,\varphi)=\frac{1}{\sqrt{rR}}\psi(r,R,\varphi)
\end{equation}
and then multiply the resulting Hamiltonian by $\sqrt{rR}$. This leads to:
\begin{equation}\label{ham4}
	\begin{aligned}
		H&=-\frac{1}{4\mu}\left(\frac{\partial^2}{\partial r^2}+\frac{1}{4r^2}\right)
		-\frac{1}{\mathcal{M}}\left(\frac{\partial^2}{\partial R^2}+\frac{1}{4R^2}\right) \\
		&-\left(\frac{1}{4\mu r^2}+\frac{1}{\mathcal{M} R^2}\right)\frac{\partial^2}{\partial \varphi^2}
		+V_1(r)+V_2(R)\\
		&+W(r,R)+V_{\textmd{int}}(r),
	\end{aligned}
\end{equation}
The function $\psi(r,R,\varphi)$ behaves as $\sqrt{rR}$ at the origin. To make it behave linearly, we transform the spatial variables as:
\begin{equation}\label{xy}
	\begin{cases}
	r=\rho_{\textmd{m}}x^2 \\
	R=\rho_{\textmd{m}}y^2
	\end{cases},
\end{equation}
where $\rho_{\textmd{m}}$ is the maximum value of the spatial grid. With the transformation \eqref{xy}, the linear behavior of the wave function $\psi(x,y,\varphi)\rightarrow x$ and $\psi(x,y,\varphi)\rightarrow y$ at $x,y\rightarrow 0$ can be approximated well by finite differences \cite{melezhik_pc}.

The Hamiltonian in new variables takes the form:
\begin{equation}\label{ham5}
	\begin{aligned}
		H&=h_1(x)+h_2(y)+W(x,y)+V_{\textmd{int}}(x) \\
		&-\frac{1}{\rho_\textmd{m}^2}\left(\frac{1}{4\mu x^4}+\frac{1}{\mathcal{M} y^4}\right)\frac{\partial^2}{\partial \varphi^2}
	\end{aligned}
\end{equation}
where
\begin{equation}
	h_1(x)=-\frac{1}{16\mu\rho_\textmd{m}^2 x^2}\left(\frac{\partial^2}{\partial x^2}-\frac{1}{x}\frac{\partial}{\partial x}+\frac{1}{x^2}\right)+V_1(x),
\end{equation}
\begin{equation}
	h_2(y)=-\frac{1}{4\mathcal{M}\rho_\textmd{m}^2 y^2}\left(\frac{\partial^2}{\partial y^2}-\frac{1}{y}\frac{\partial}{\partial y}+\frac{1}{y^2}\right)+V_2(y),
\end{equation}
\begin{equation}
	V_1(x)=\mu^2\left(\frac{\omega_1^2}{m_1}+\frac{\omega_2^2}{m_2}\right)\rho_{\textmd{m}}^2x^4,
\end{equation}
\begin{equation}
	V_2(y)=\frac{1}{4}\left(m_1\omega_1^2+m_2\omega_2^2\right)\rho_{\textmd{m}}^2y^4,
\end{equation}
\begin{equation}\label{int}
	V_{\textmd{int}}(x)=V_0\exp\left\{ -\dfrac{\rho_{\textmd{m}}^2}{\rho_0^2}x^4 \right\}
\end{equation}
\begin{equation}
	W(x,y)=\mu\left(\omega_1^2-\omega_2^2\right)\rho_{\textmd{m}}^2x^2y^2\cos\varphi.
\end{equation}

We expand the sought wave function in terms of functions $\xi_m(\varphi)$ \cite{koval,melezhik} as:
\begin{equation}\label{wf}
	\begin{aligned}
	\psi(x,y,\varphi)&=\sum\limits_{j=0}^{2M}\sum\limits_{m=-M}^{M}\xi_m(\varphi)\xi^{-1}_{mj}\psi_j(x,y) \\
	&=\frac{1}{2M+1}\sum\limits_{j=0}^{2M}\sum\limits_{m=-M}^{M}e^{im(\varphi-\varphi_j)}\psi_j(x,y),
	\end{aligned},
\end{equation}
where
\begin{equation}
	\xi_{jm}=\frac{(-1)^m}{\sqrt{2\pi}}e^{im\varphi_j}
\end{equation}
\begin{equation}
	\xi^{-1}_{m j}=\frac{\sqrt{2\pi}(-1)^m}{2M+1}e^{-im\varphi_j}
\end{equation}
\begin{equation}
	\varphi_j=\frac{2\pi j}{2M+1},\hspace{.5cm} j=0,1,2,...,2M
\end{equation}
Thus, the Hamiltonian in matrix form can be represented as:
\begin{equation}
	\begin{aligned}
	H_{jj'}(x,y)&=\bigg(h_1(x)+V_{\textmd{int}}(x)+h_2(y)+W_j(x,y)\bigg)\delta_{jj'} \\
	&+\Omega_{jj'}(x,y),
\end{aligned}
\end{equation}
where
\begin{equation}
	\begin{aligned}
	\Omega_{jj'}&=
	\frac{1}{\rho_\textmd{m}^2}\left(\frac{1}{4\mu x^4}+\frac{1}{\mathcal{M}y^4}\right)
	\sum\limits_{m=-M}^{M}m^2\xi_{jm}\xi^{-1}_{m j'} \\
	&=
	\frac{1}{\rho_\textmd{m}^2(2M+1)}\left(\frac{1}{4\mu x^4}+\frac{1}{\mathcal{M}y^4}\right)
	\sum\limits_{m=-M}^{M}m^2e^{im(\varphi_j-\varphi_{j'})}
	\end{aligned}
\end{equation}

To investigate the quench dynamics, we numerically solve the time-dependent Schr\"{o}dinger equation given by:
\begin{equation}\label{tdse}
	i\frac{\partial}{\partial t}\psi(t)=H\psi(t),
\end{equation}
where we omit the dependence on $x,y,\varphi$ when not necessary, and $H$ is the Hamiltonian derived earlier. To tackle \eqref{tdse}, we employ the split-operator method to approximate the evolution operator, resulting in the following time-stepping scheme:
\begin{equation}\label{split}
	\begin{aligned}
	\psi&(t+\Delta t)=\exp\left(-\frac{i}{2}\Delta t \big[\hat{\Omega}+\hat{W}\big]\right) \\
	&\times\exp\big(-i\Delta t\big[\hat{h}(x)+V_{\textmd{int}}(x)\big]\big)\\
	&\times\exp\big(-i\Delta t\hat{h}(y)\big)\exp\left(-\frac{i}{2}\Delta t\big[\hat{\Omega}+\hat{W}\big]\right)\psi(t)
	\end{aligned}
\end{equation}
The action of the operators in \eqref{split} is approximated using the Crank-Nicolson formula as follows:
\begin{equation}\label{cn}
	\exp\left(-\Delta t \hat{A}\right)=\left(1+\frac{i}{2}\Delta t \hat{A}\right)^{-1}\left(1-\frac{i}{2}\Delta t \hat{A}\right),
\end{equation}
where $\hat{A}$ represents the operators in \eqref{split}. The operators containing spatial derivatives are approximated using central differences with sixth-order accuracy. Both \eqref{split} and \eqref{cn} exhibit an accuracy order of $\mathcal{O}(\Delta t^3)$. In our calculations, we set $\rho_\textmd{m}=10$, and use 50 grid points for each variable - $x$ and $y$, and 11 grid points for the variable $\varphi$. For the time step, we use $\Delta t=0.001$. These grid point numbers and the time step ensure a satisfactory convergence of the results. Different implementations of \eqref{split} and \eqref{cn} were explored in \cite{shadmehri, ishmukh6}.

We use the following values for the particle masses and for the frequencies in the case of the anisotropic trap:
\begin{equation}\label{parameters}
	\begin{aligned}
		&m_1=1,\\
		&m_2=0.5,\\
		&\omega_1=0.8,\\
		&\omega_2=1.7.
	\end{aligned}
\end{equation}
While this is not a unique set of parameters, there are other values that can be used to detect the molecular excited state we are in search of. However, we have found that when using the values given in \eqref{parameters}, the emergence of the molecular excited state is quite pronounced.

To compute the bound states, we utilize the imaginary time propagation technique. For calculating the excited states, we employ the Gram-Schmidt procedure to orthogonalize each state

One of our primary quantities of interest for analyzing the quench dynamics is the overlap integral $\mathcal{Q}_{n,N}(t)$, defined as follows:
\begin{equation}
	\begin{aligned}
	&\mathcal{Q}_{n,N}(t)\\
	&=8\pi\rho_m^2\int\limits_{0}^{2\pi}\int\limits_{0}^{1}\int\limits_{0}^{1} d\varphi dxdy\cdot xy \psi_{n,N}(x,y,\varphi)\psi(x,y,\varphi,t)
	\end{aligned},
\end{equation}
where the indices $(n,N)$ refer to the quantum numbers associated with the relative and center-of-mass motion excited states, respectively. Here, $\psi_{n,N}(x,y,\varphi)$ represents the stationary wave function. When $\psi_{n,N}(x,y,\varphi)$ corresponds to the initial state, the quantity is known as the fidelity, denoted as $F(t)$.

\section{Spectrum}\label{spectrum}

\begin{figure}%[H]
	\centering
	\includegraphics[width=9cm,clip]{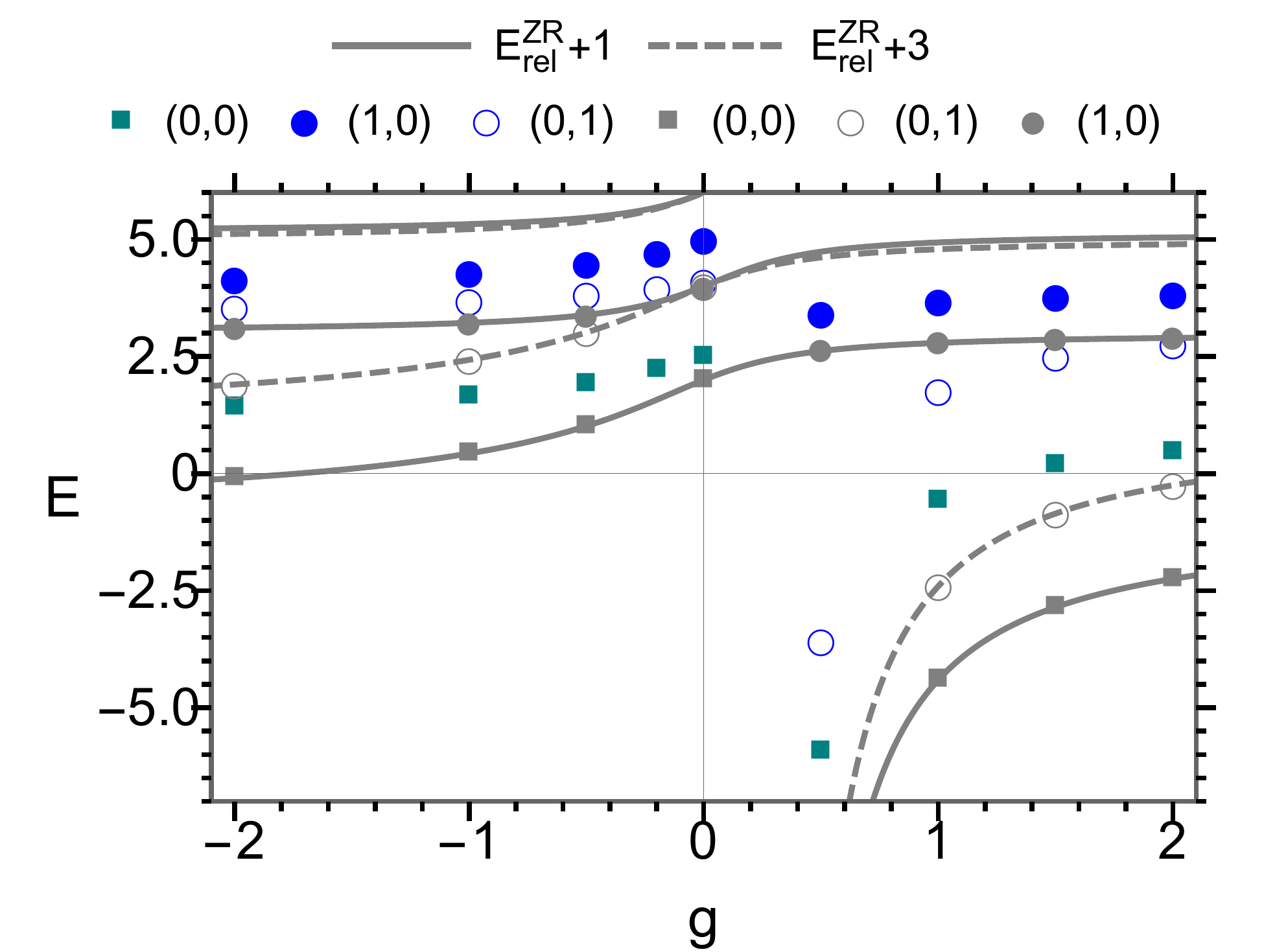}
	\caption{(Color online) Bound energy levels for the ground and excited states in both the isotropic (gray points) and anisotropic (color points) traps as functions of the coupling strength $g$. The indices $(n,N)$ refer to the quantum numbers of the relative and center-of-mass motions. The gray solid and dashed lines correspond to the analytical results from \cite{bougas}.}\label{energy}
\end{figure}

In Fig.~\ref{energy}, we present the energy levels of the three lowest bosonic states for both isotropic (gray points) and anisotropic (color points) traps, along with the analytically calculated levels from \cite{bougas} (represented by solid and dashed lines), as functions of the coupling strength of the contact interaction, denoted as $g$. In the case of the isotropic harmonic trap, where the two particles have equal masses and frequencies, the relative and center-of-mass motions are decoupled. This case, for the zero-range interaction, has been studied in \cite{bougas}. To ensure a valid comparison with the results presented in that study, we adjust the interaction strength of the potential $V_{\textmd{int}}$ to obtain the same energy level for the ground state as the one obtained using the pseudopotential in \cite{bougas}. In our notation, we refer to the ground state as the state with no nodes in the $xy$ plane, whereas states with nodes along the $x$ or $y$ directions are termed relative or center-of-mass excited states, labeled with indices $(n,N)$, respectively. This notation differs from that used in \cite{bougas}. From Fig.~\ref{energy}, it is evident that our finite-range interaction calculation yields results that are almost indistinguishable from the analytical results, even for excited states. Thus, our finite-range interaction can be considered to some extent as a valid approximation of the zero-range interaction. In the case of the anisotropic trap, the energy levels are higher compared to those in the isotropic trap. This difference is due to the anisotropic trap's narrower size compared to the isotropic harmonic trap. At $g=0$, the degeneracy of the energy level is lifted due to the presence of anisotropic terms.

In Fig.~\ref{wf}, we present the probability densities integrated over the angular variable for $g=-1$ and $g=1$ for the three lowest stationary states in the case of the anisotropic trap. We focus on analyzing the impacts from these states only, as their contributions are expected to be predominant among the other states \cite{bougas,ishmukh3}.

\begin{figure}
	\centering
	\textbf{$g=-1$}\par\medskip
	\begin{tabular}{ccc}
		$(0,0)$ & $(0,1)$ & $(1,0)$ \\
		\includegraphics[width=.15\textwidth]{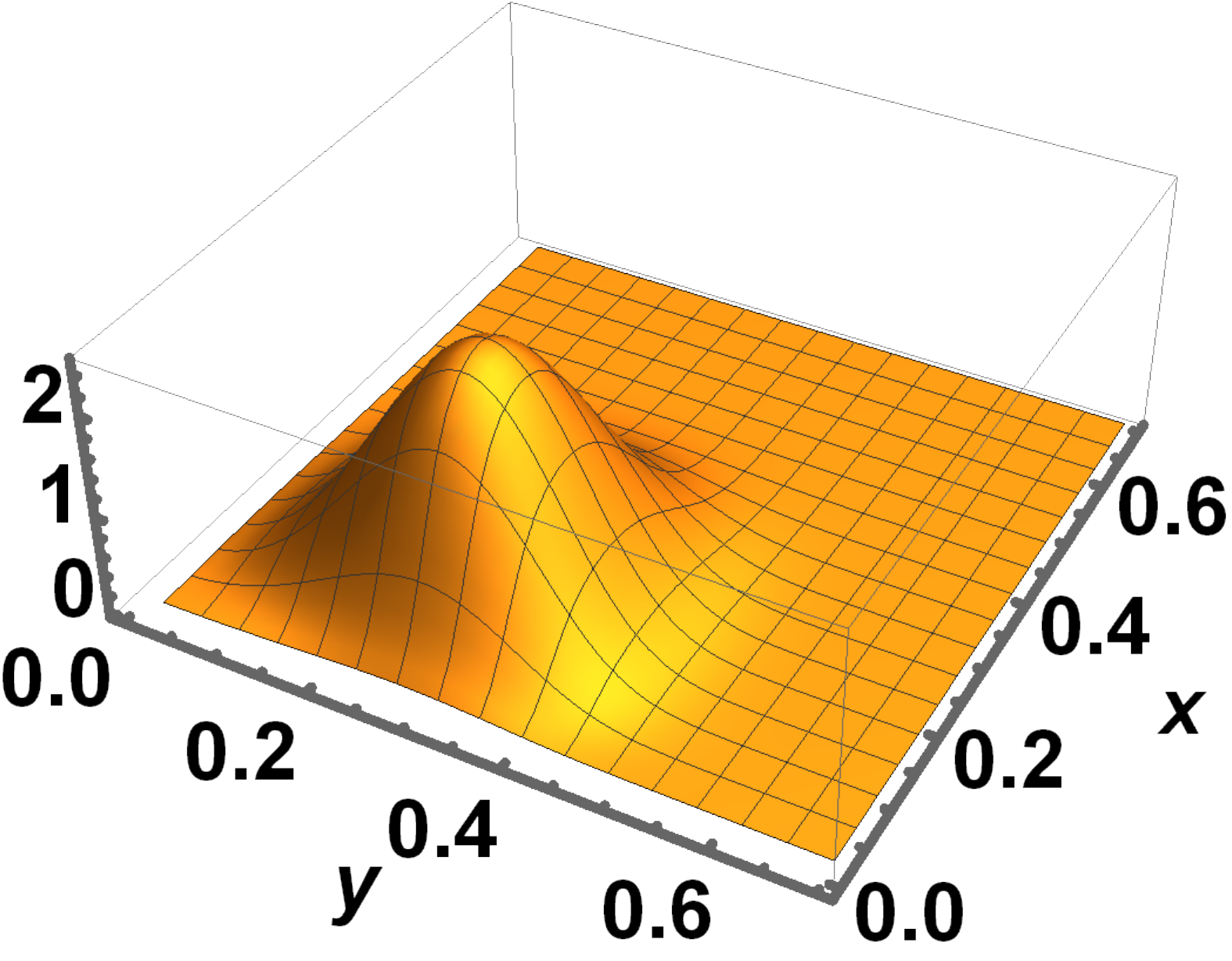} &
		\includegraphics[width=.15\textwidth]{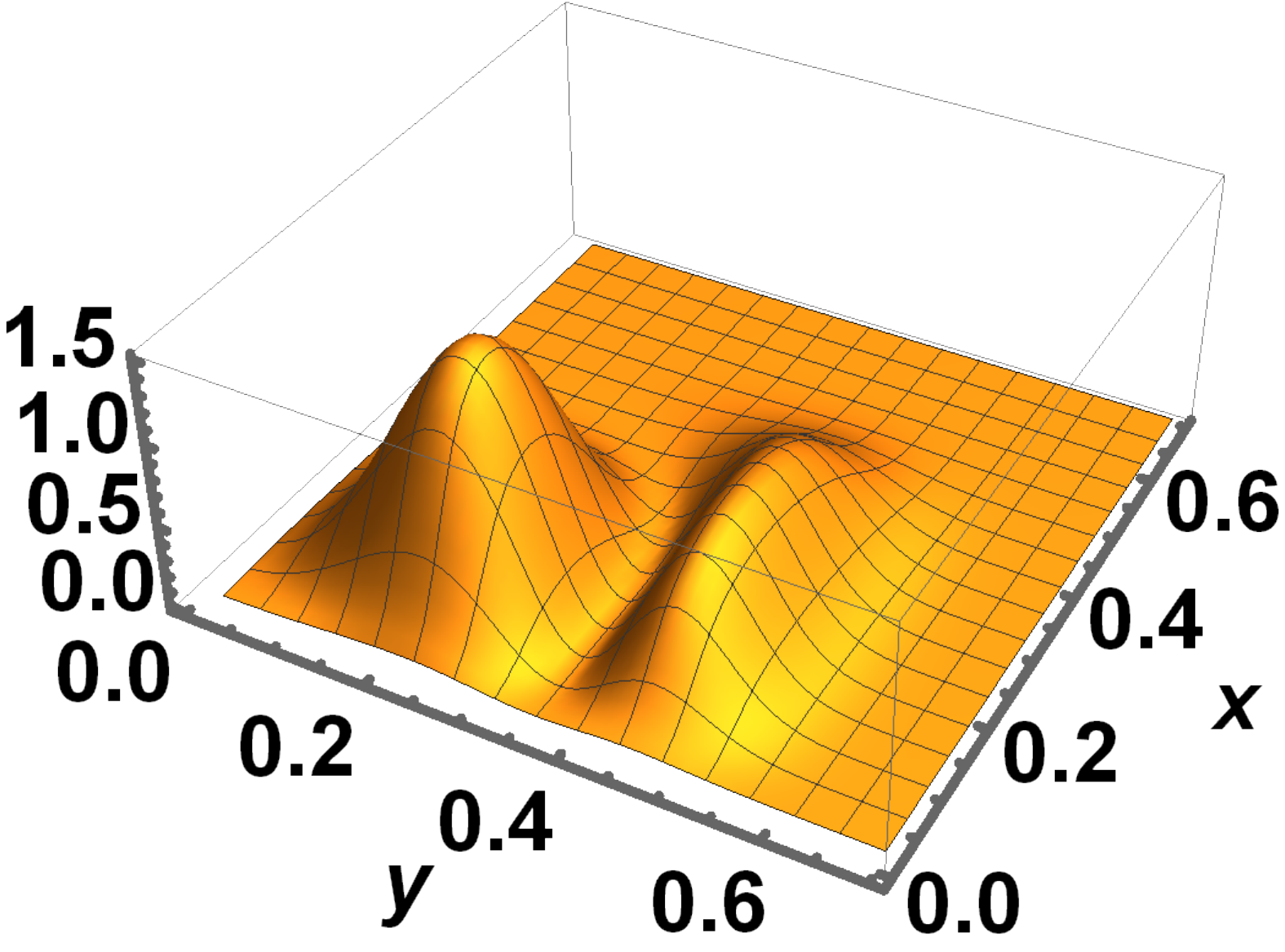} &
		\includegraphics[width=.15\textwidth]{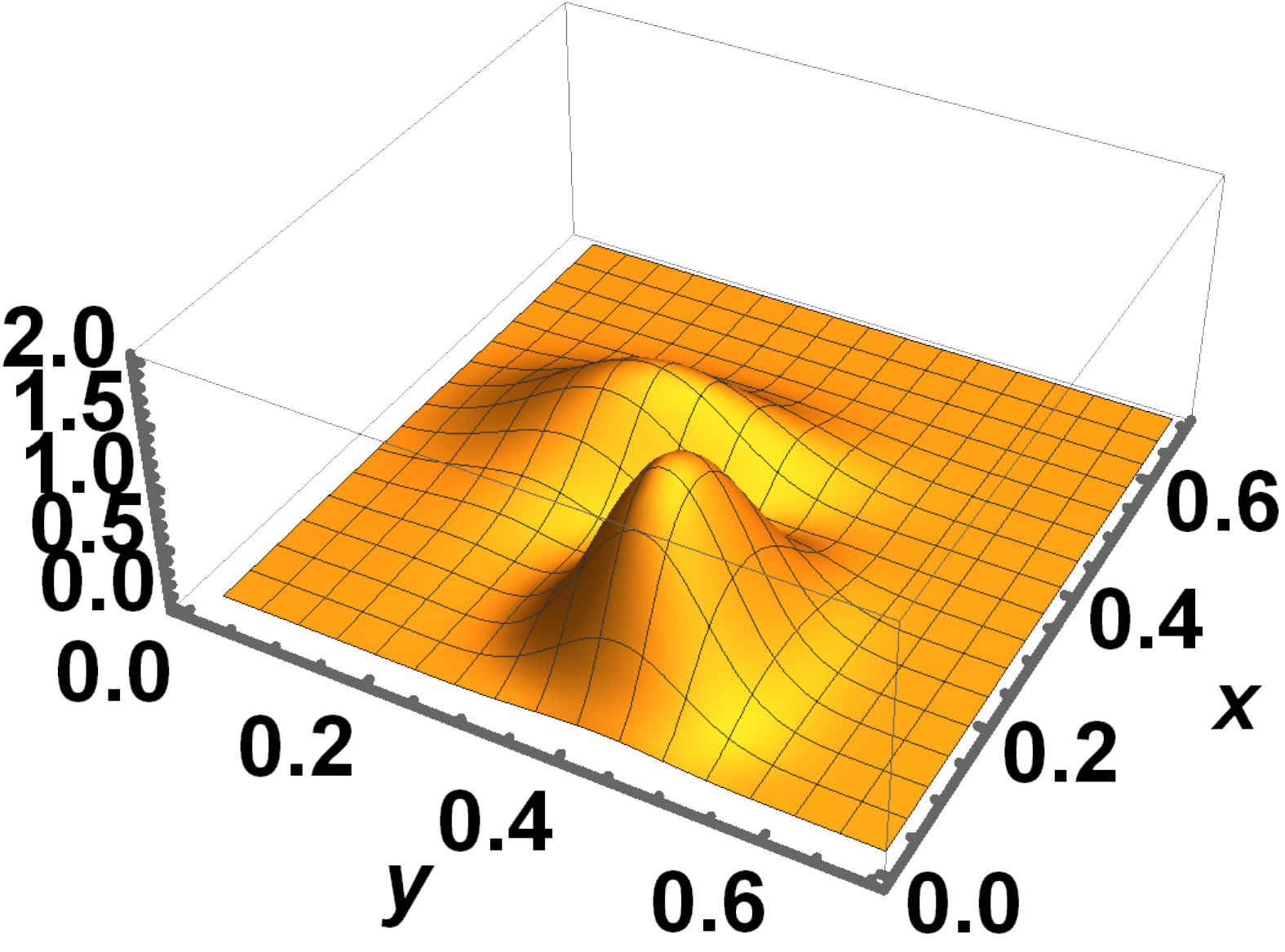} \\
	\end{tabular}
	\textbf{$g=1$}\par\medskip
	\begin{tabular}{ccc}
		$(0,0)$ & $(0,1)$ & $(1,0)$ \\
	\includegraphics[width=.15\textwidth]{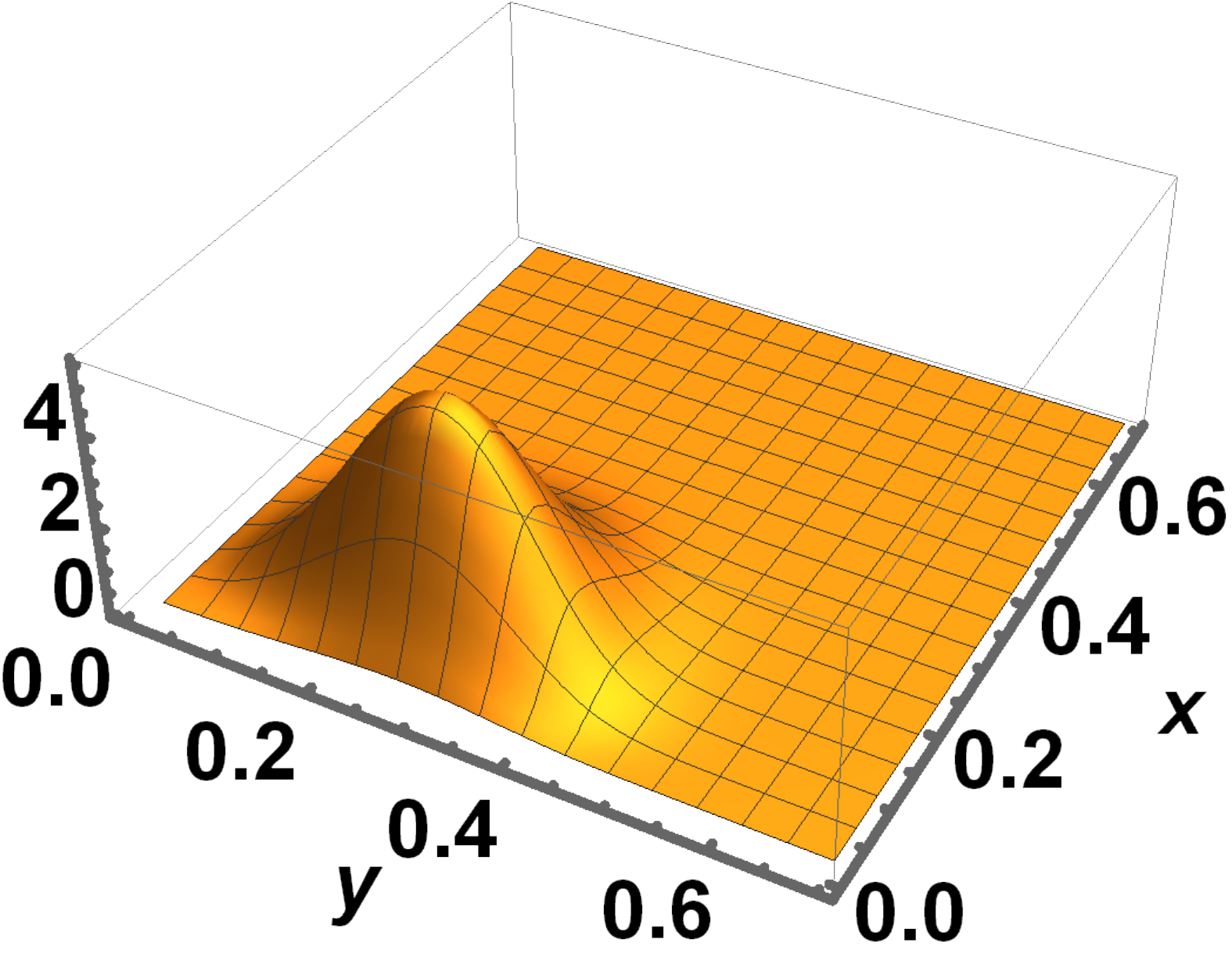} &
	\includegraphics[width=.15\textwidth]{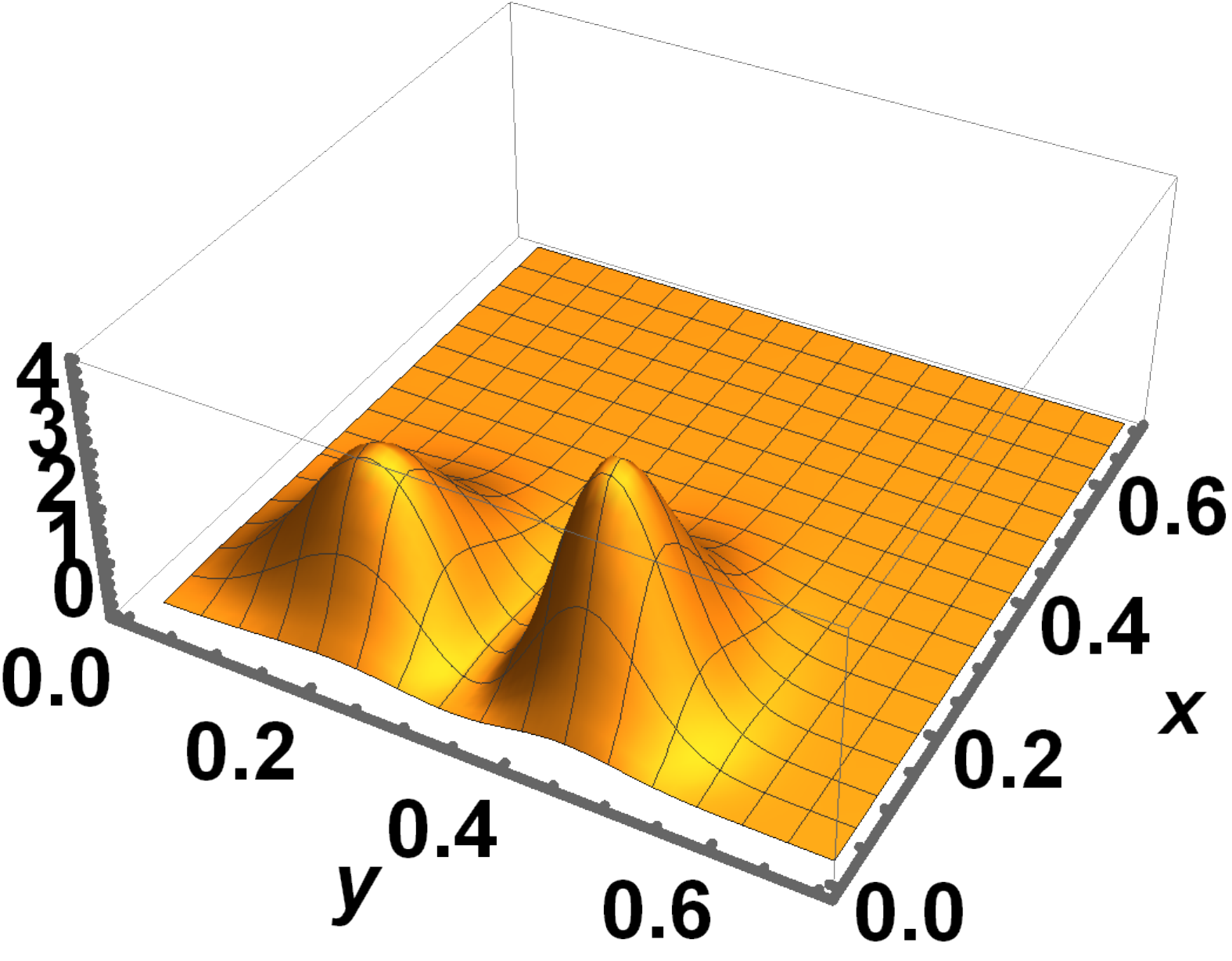} &
	\includegraphics[width=.15\textwidth]{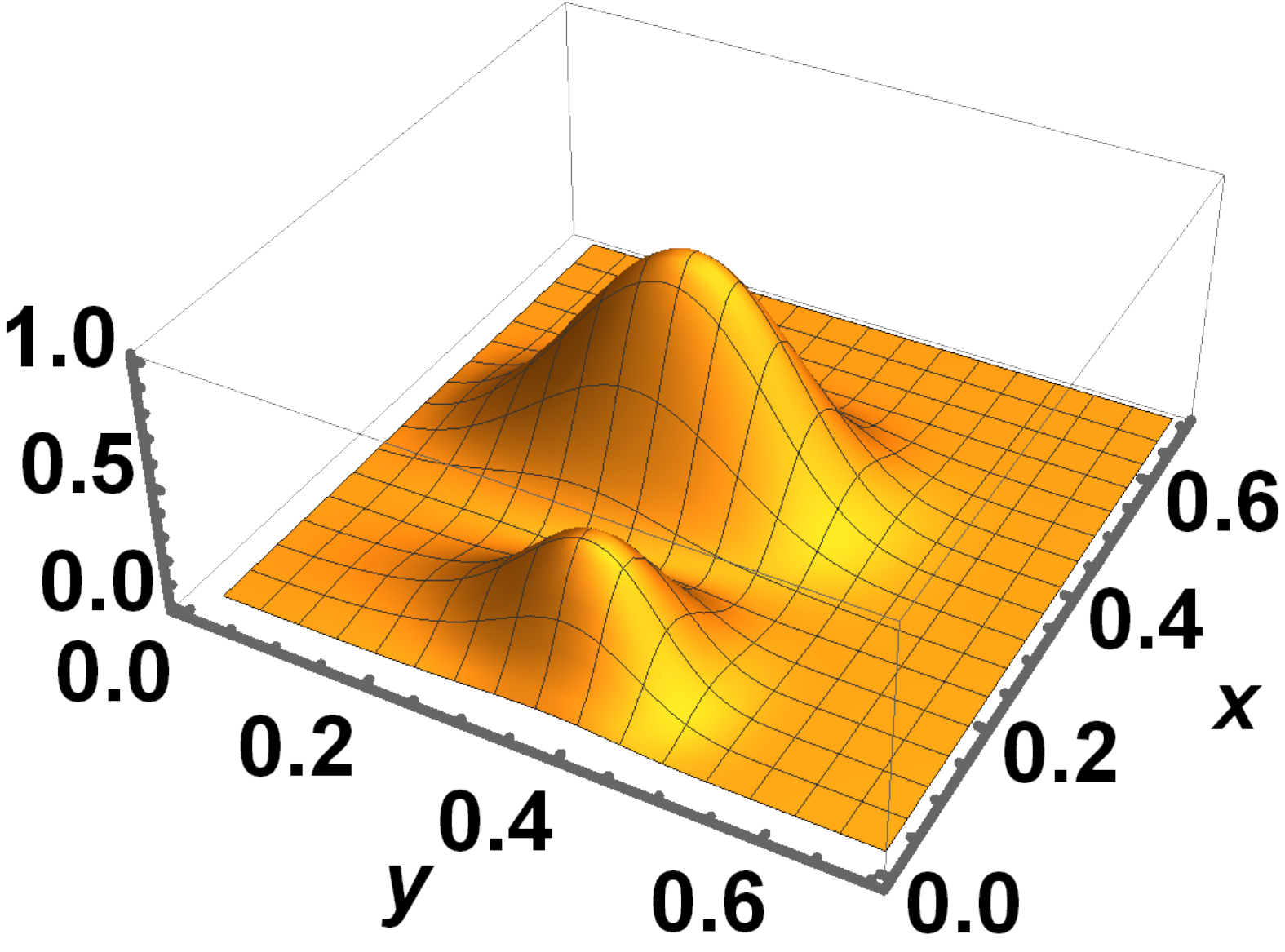} \\
	\end{tabular}
	\caption{(Color online) Stationary wave functions $\int d\varphi|\psi(x,y,\varphi, t=0)|^2$ integrated over the angular variable $\varphi$. Indices $(n,N)$ refer to the nodes of the wave function.}\label{wf}
\end{figure}

\section{Isotropic harmonic trap}\label{isotropic}

\begin{figure}%[H]
	\centering
	\includegraphics[width=8.5cm,clip]{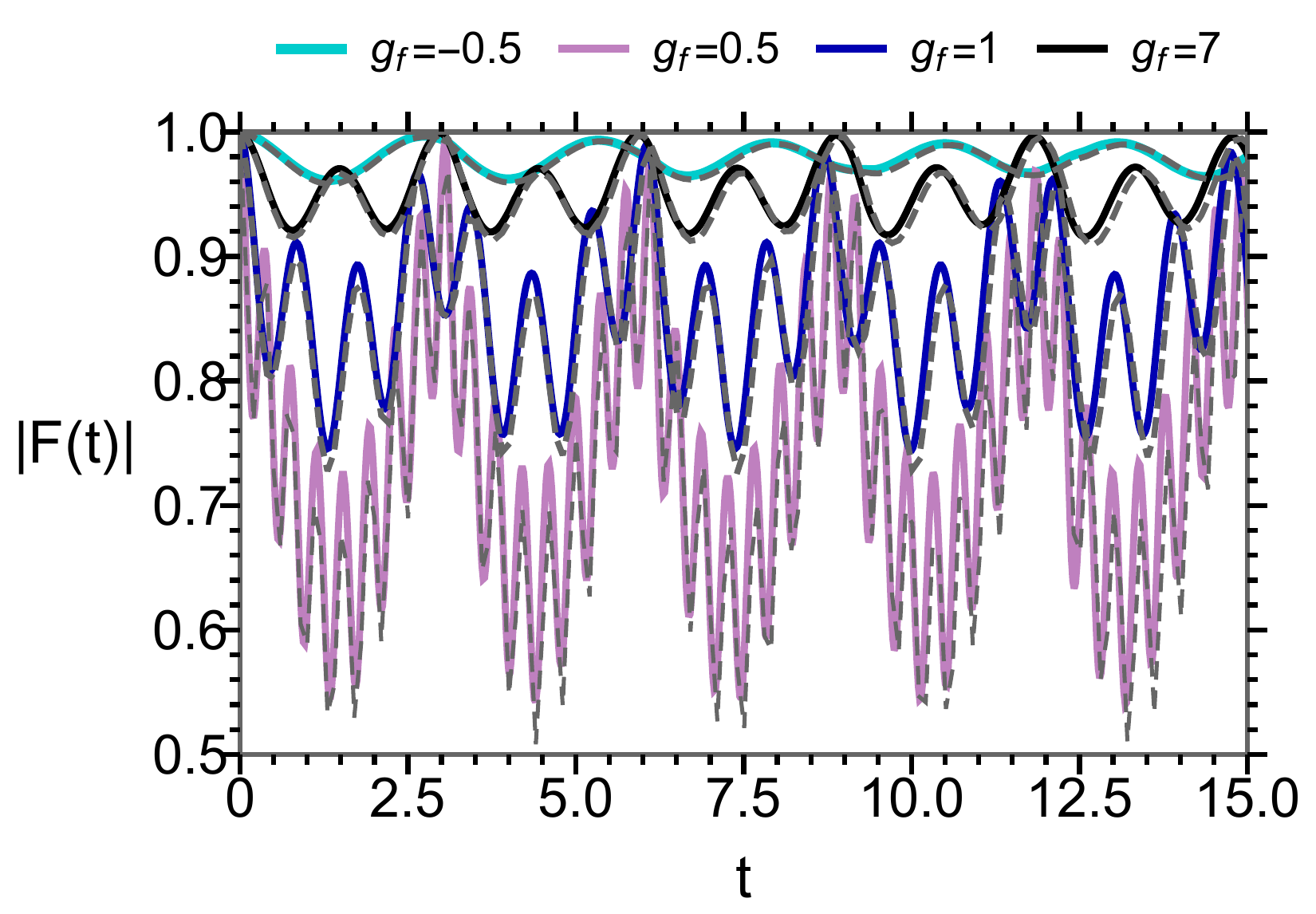}
	\caption{(Color online) Fidelity $F(t)$ for different values of the coupling constant $g_{\textmd{f}}$ in the case of the isotropic harmonic trap. The initial coupling constant is $g_{\textmd{in}}=-1$. The solid lines refer to our numerical calculation for the Gaussian interaction potential, and the dashed lines represent the analytical results for the zero-range interaction \cite{bougas_data}.}\label{fwd_ho}
\end{figure}

\begin{figure}%[H]
	\centering
	\includegraphics[width=8.5cm,clip]{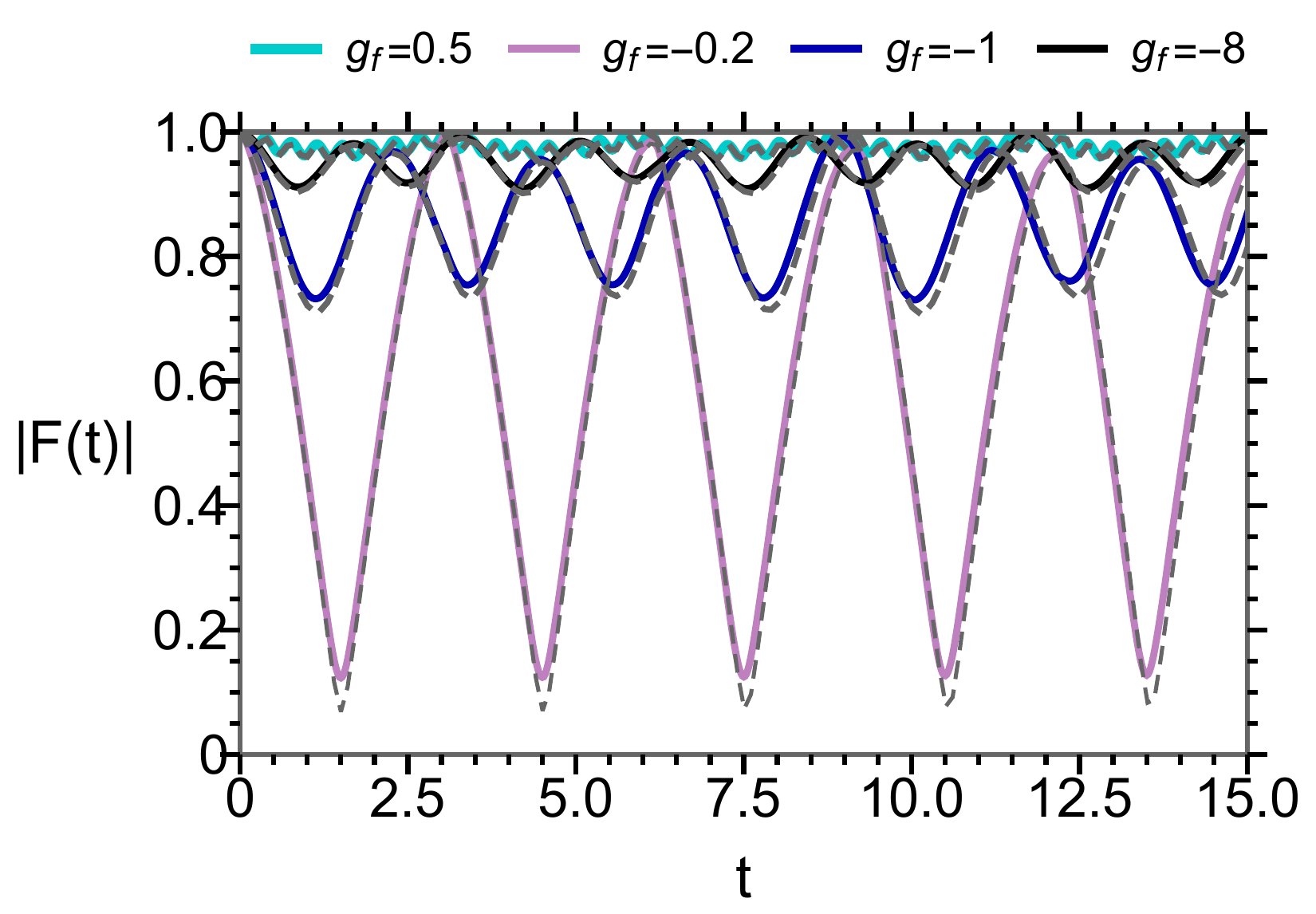}
	\caption{(Color online) Fidelity $F(t)$ for different values of the coupling constant $g_{\textmd{f}}$ in the case of the isotropic harmonic trap. The initial coupling constant is $g_{\textmd{in}}=1$. The solid lines refer to our numerical calculation for the Gaussian interaction potential, and the dashed lines represent the analytical results for the zero-range interaction \cite{bougas_data}.}\label{bwd_ho}
\end{figure}

In this section we examine the quench dynamics in isotropic harmonic trap, starting from $g_{\textmd{in}}=-1$ (Fig.~\ref{fwd_ho}) and $g_{\textmd{in}}=1$ (Fig.~\ref{bwd_ho}). The masses of the particles and the trap frequencies for the each particle are equal to one. Throughout the paper, we consider the interaction quench starting from the relative excited state $(1,0)$, as this state is driven more efficiently out of equilibrium \cite{bougas}.  The oscillation patterns in these cases have been extensively analyzed in \cite{bougas}, so we will not dwell on discussing them further. Instead, we compare our finite-range calculation with the zero-range calculation conducted in \cite{bougas}. Both figures demonstrate a good agreement between these two approaches. While some deviations in the fidelities between the two interaction potentials persist, they do not alter the qualitative behavior. 

\begin{figure}%[H]
%	\centering
	\begin{tabular}{cc}
		$g=-1$ & $g=1$ \\
		\includegraphics[width=.25\textwidth]{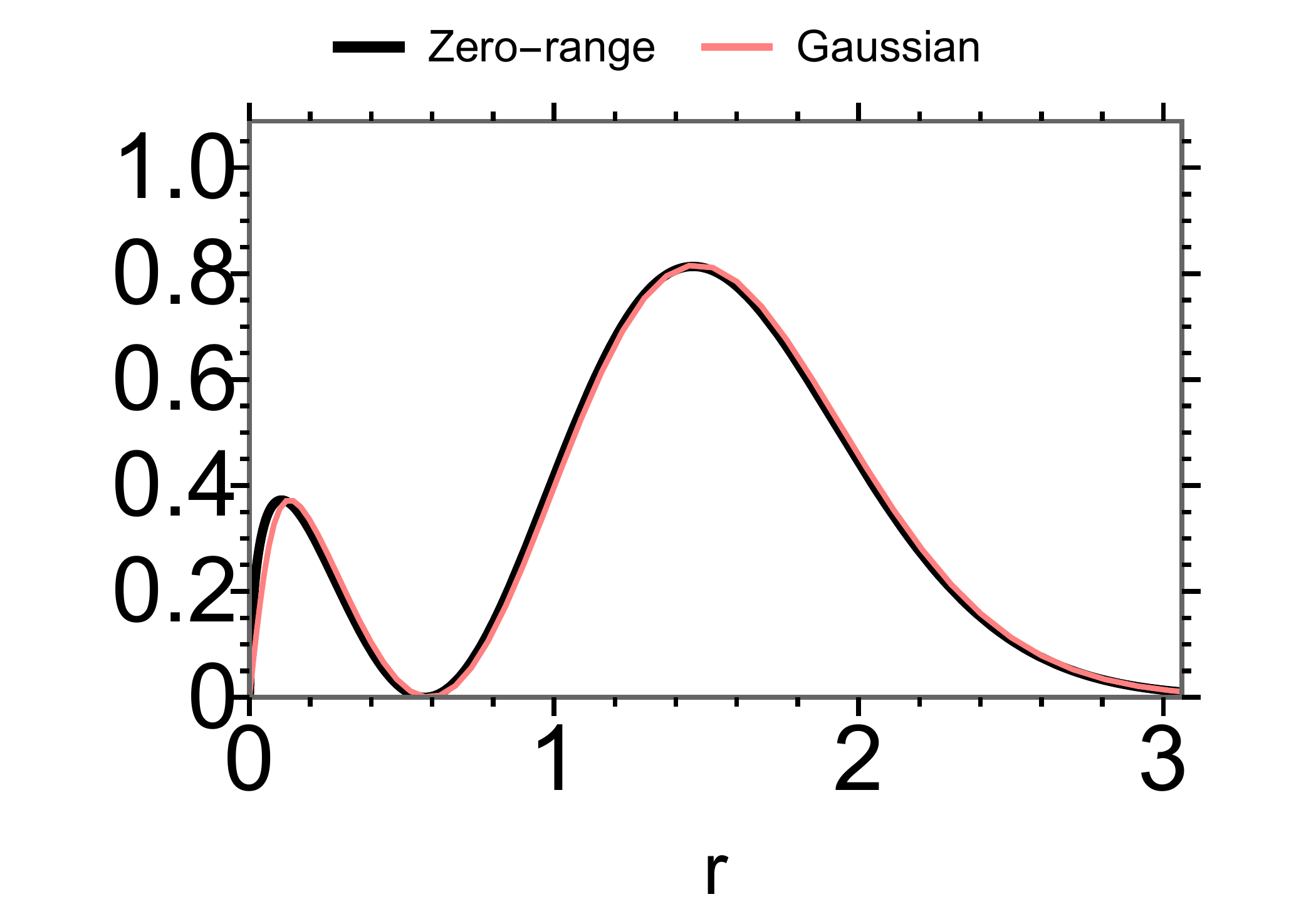} &
		\includegraphics[width=.25\textwidth]{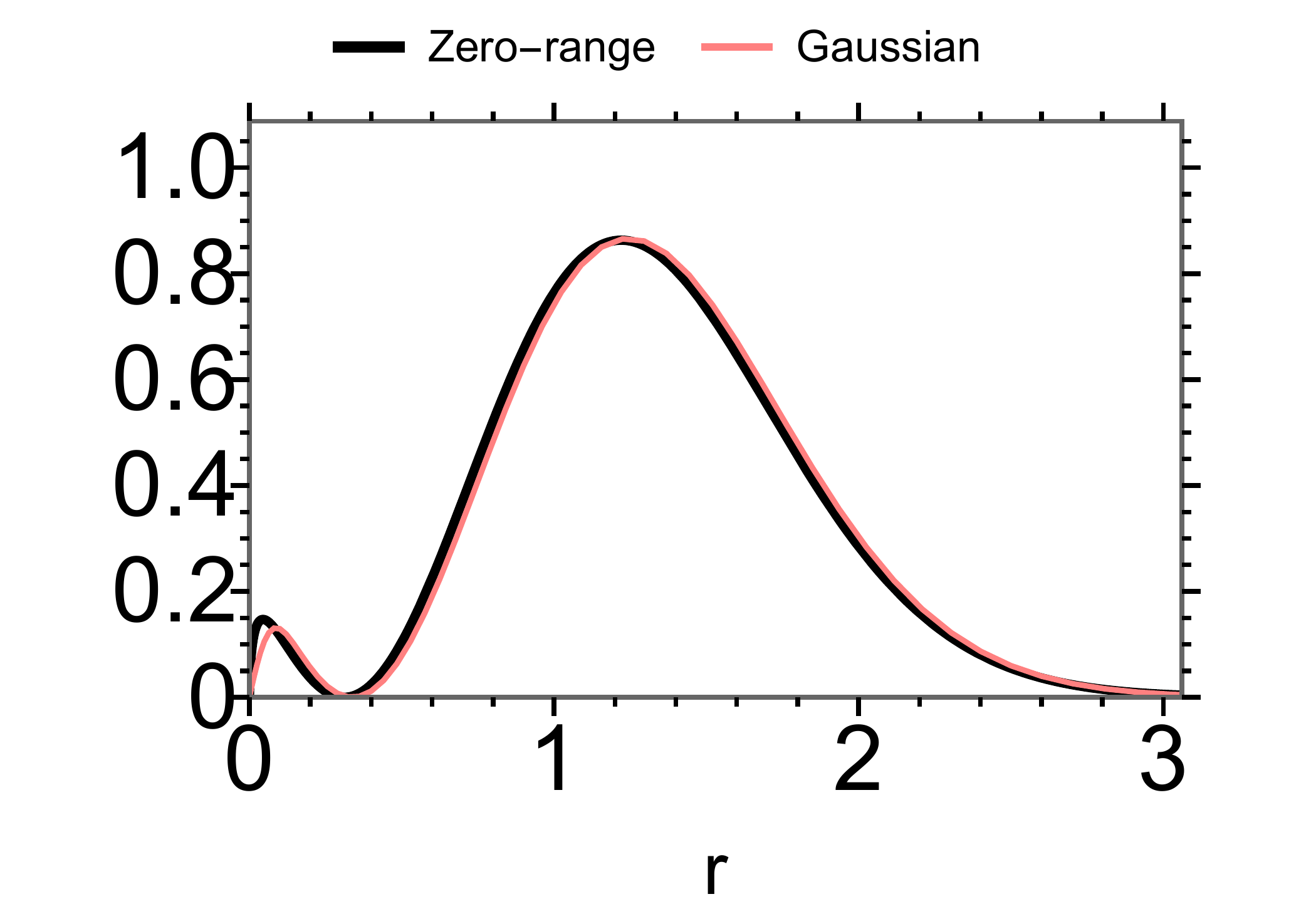} \\
	\end{tabular}
	\caption{(Color online) Comparison between the zero-range interaction and Gaussian interaction potentials for the radial probability density $2\pi r |\psi(r)|^2$ of the first excited state.}\label{1d_wf}
\end{figure}

Upon analyzing the radial probability density for the relative motion (Fig.~\ref{1d_wf}) for both types of interactions, a noticeable discrepancy emerges primarily in the region around the width of the Gaussian interaction ($r\simeq\rho_0=0.1$). These small differences in the wave functions might be responsible for the slight deviations and phase lag that we observe in the fidelity between the two models, as they involve the overlap of the final wave functions with the initial one \cite{bougas_pc}.

Thus, our results for the finite-range interaction potential can be interpreted in terms of the coupling strength of the zero-range interaction potential to a good accuracy.

\section{Anisotropic harmonic trap}\label{anisotropic}
\subsection{Quench dynamics from the attractive to the repulsive interaction}

\begin{figure}%[H]
	\centering
	\includegraphics[width=8.5cm,clip]{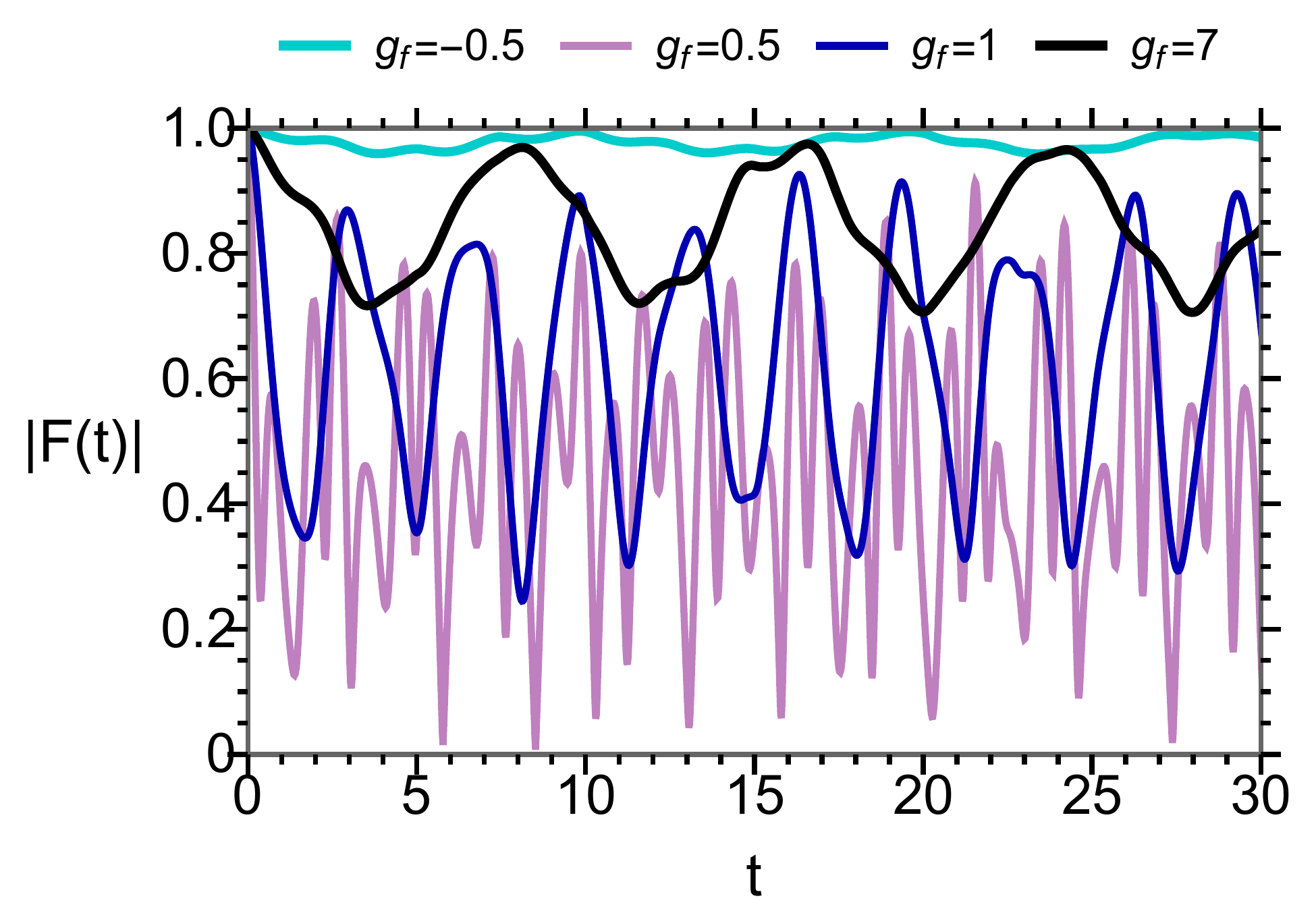}
	\caption{(Color online) Fidelity $F(t)$ for different values of the coupling constant $g_{\textmd{f}}$ in the case of the anisotropic harmonic trap. The initial coupling constant is $g_{\textmd{in}}=-1$.}\label{fwd_2}
\end{figure}

In the case of the anisotropic trap, we consider two atoms with different masses confined in a trap with varying frequencies, as specified in \eqref{parameters}. The coupling constant of the initial state is set to $g_{\textmd{in}}=-1$, and at $t>0$, we change it to different values of $g_{\textmd{f}}$. Fig.~\ref{fwd_2} displays the fidelity for such quench dynamics. The values of $g_{\textmd{f}}$ used are the same as in the case of the isotropic trap examined earlier. By comparing the isotropic and anisotropic trap cases, we observe distinct oscillation behaviors, but the overall tendency remains similar. That is, at $g_{\textmd{f}}=-0.5$, there is minimal deviation of the fidelity $|F(t)|$ from unity, and as $g_{\textmd{f}}$ increases, the oscillation amplitude rises, as observed for $g_{\textmd{f}}=0.5$. However, at a certain point, the oscillation amplitude starts decreasing, which can be observed for $g_{\textmd{f}}=1$ and $g_{\textmd{f}}=7$. Additionally, we note that the oscillation behavior in the anisotropic case is more distorted and less uniform compared to the isotropic case. The oscillation amplitude and period are significantly larger. In the isotropic trap, two distinct frequencies, i.e. slower and faster ones, can be distinguished for $g_{\textmd{f}}=0.5$ and $g_{\textmd{f}}=1$. However, the contribution of the faster frequency is less pronounced in the anisotropic trap.

\begin{figure}%[H]
	\centering
	\includegraphics[width=8.5cm,clip]{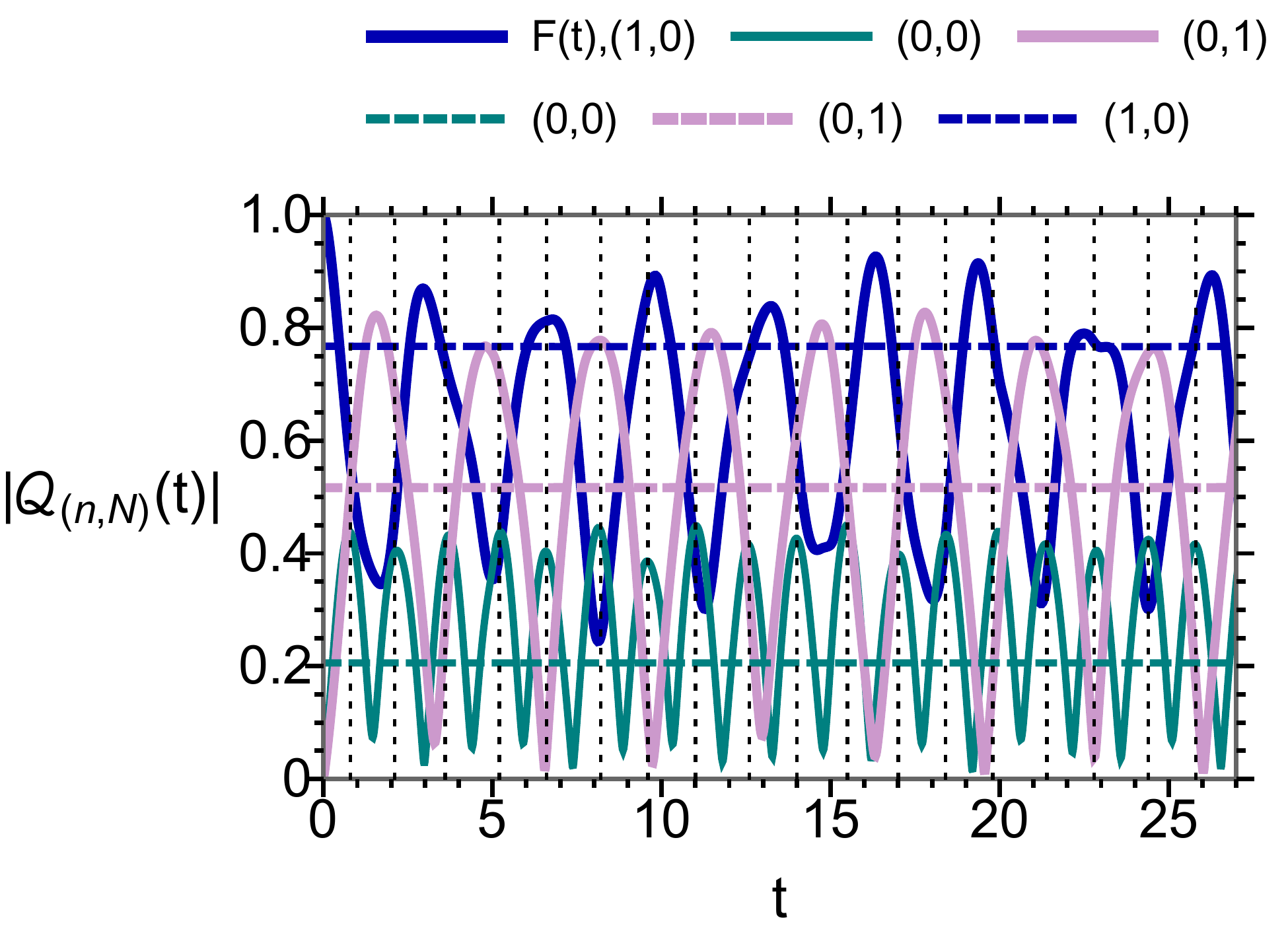}
	\caption{(Color online) Fidelity $F(t)$ and the overlap integrals $\mathcal{Q}$ between the time-evolving state $\psi(t)$ and different pre- (solid lines) and postquench (dashed lines) states in the case of the anisotropic harmonic trap. The indices $(n,N)$ refer to the states with the quantum numbers of the relative and center-of-mass motions. The system of the two atoms is prepared in the $(1,0)$ excited state with $g_{\textmd{in}}=-1$ and quenched to $g_{\textmd{f}}=1$. The dashed vertical lines indicate the emergence of the molecular excited state - $(0,1)$.}\label{fwd}
\end{figure}

Next, we consider the scenario where the coupling constant is fixed at $g_{\textmd{f}}=1$ for $t>0$, and we monitor the dynamics of the overlap integrals between the evolving state and various prequench (solid curves) and postquench (dashed lines) states, as shown in Fig.~\ref{fwd}. The overlaps with the postquench states remain constant over time due to the evolving state being a sum over orthogonal postquench states, as explained in \cite{bougas}. The fidelity $F(t)$ and the overlap with the $(0,1)$ prequench state exhibit rapid oscillations in antiphase. The oscillation frequency of the overlap with the $(0,0)$ prequench ground state is even higher, albeit with a lower amplitude. Our focus lies particularly on the emergence of the $(0,1)$ state, and we identified its occurrence by examining the evolution of the probability density. We denote these events with vertical dashed lines, which align approximately with the peaks of the overlap with the $(0,0)$ prequench state. The overlap with the $(0,1)$ postquench excited state also remains high, indicating the emergence of the center-of-mass (molecular) excited state. However, predicting the timing of the molecular excited state solely from the overlaps with the postquench states is not straightforward.

\begin{figure}%[H]
	\centering
	\begin{tabular}{ccc}
		$t=16.4$ & $t=16.6$ & $t=17.2$ \\
		\includegraphics[width=.15\textwidth]{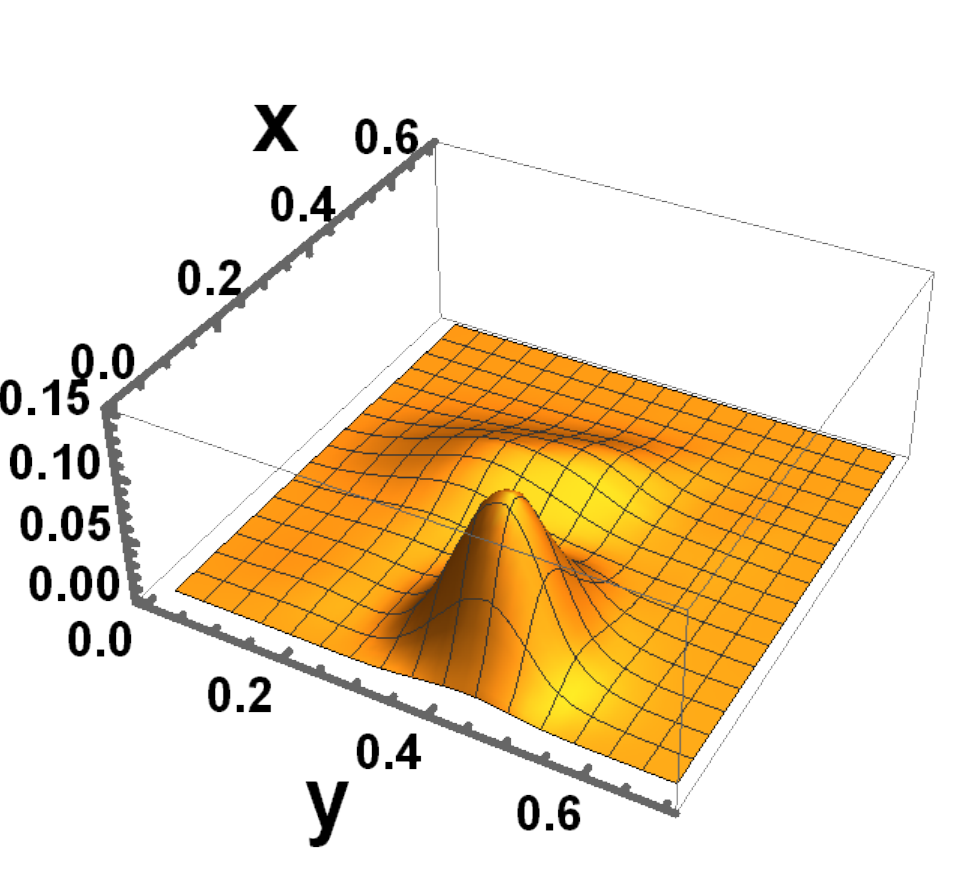} &
		\includegraphics[width=.15\textwidth]{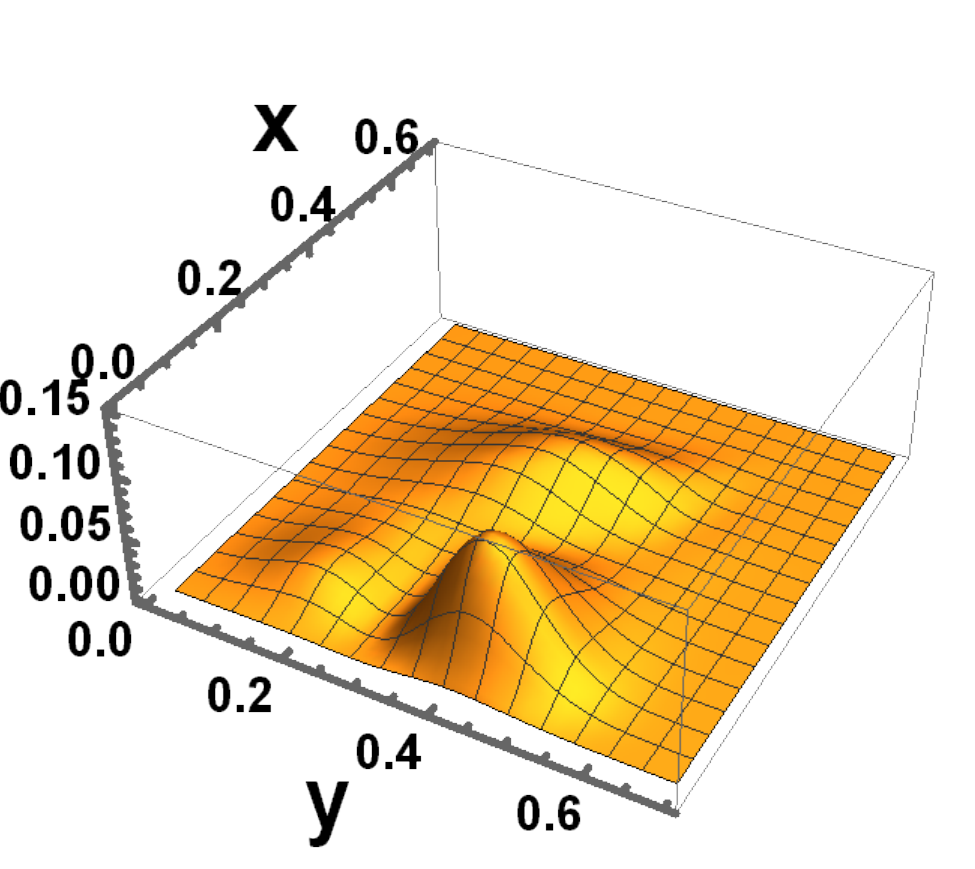} &
		\includegraphics[width=.15\textwidth]{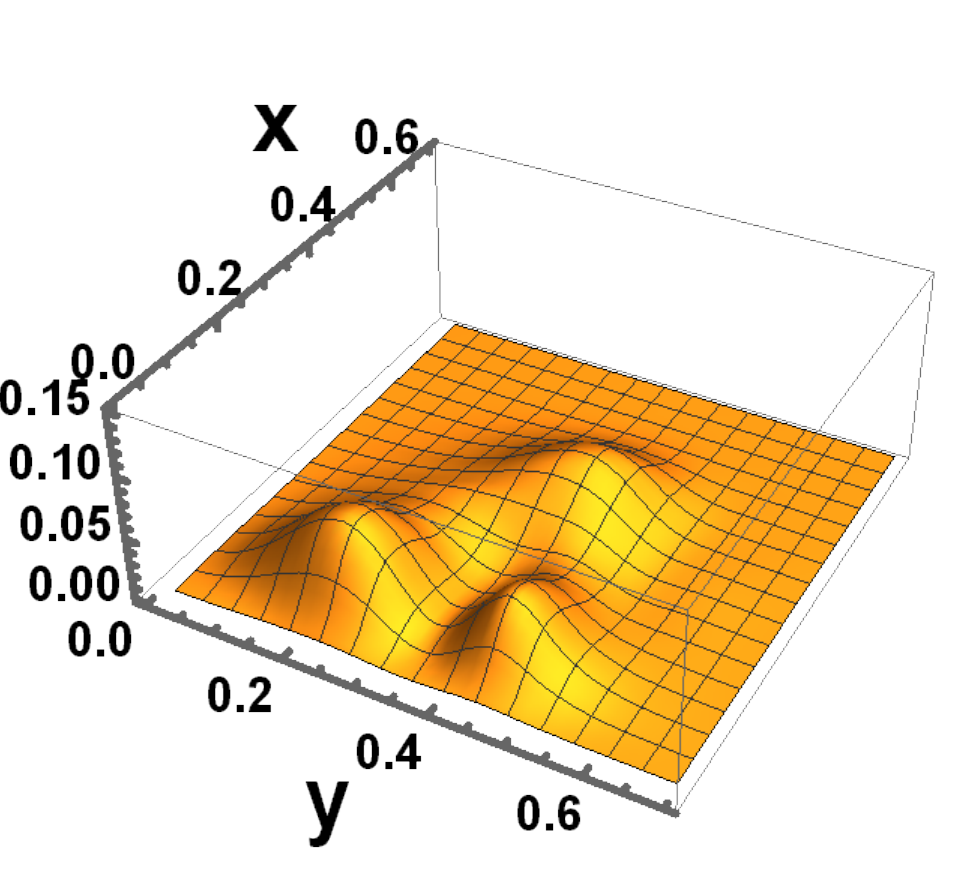} \\
		$t=17.4$ & $t=17.8$ & $t=18$\\
		\includegraphics[width=.15\textwidth]{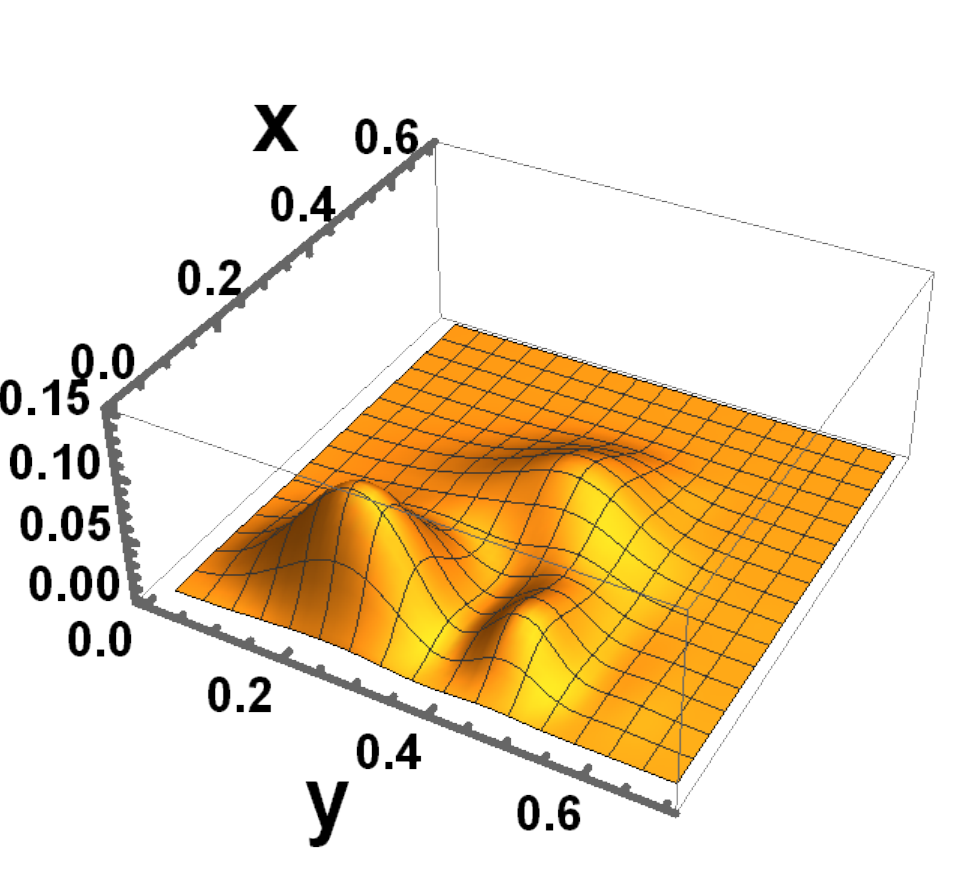} &
		\includegraphics[width=.15\textwidth]{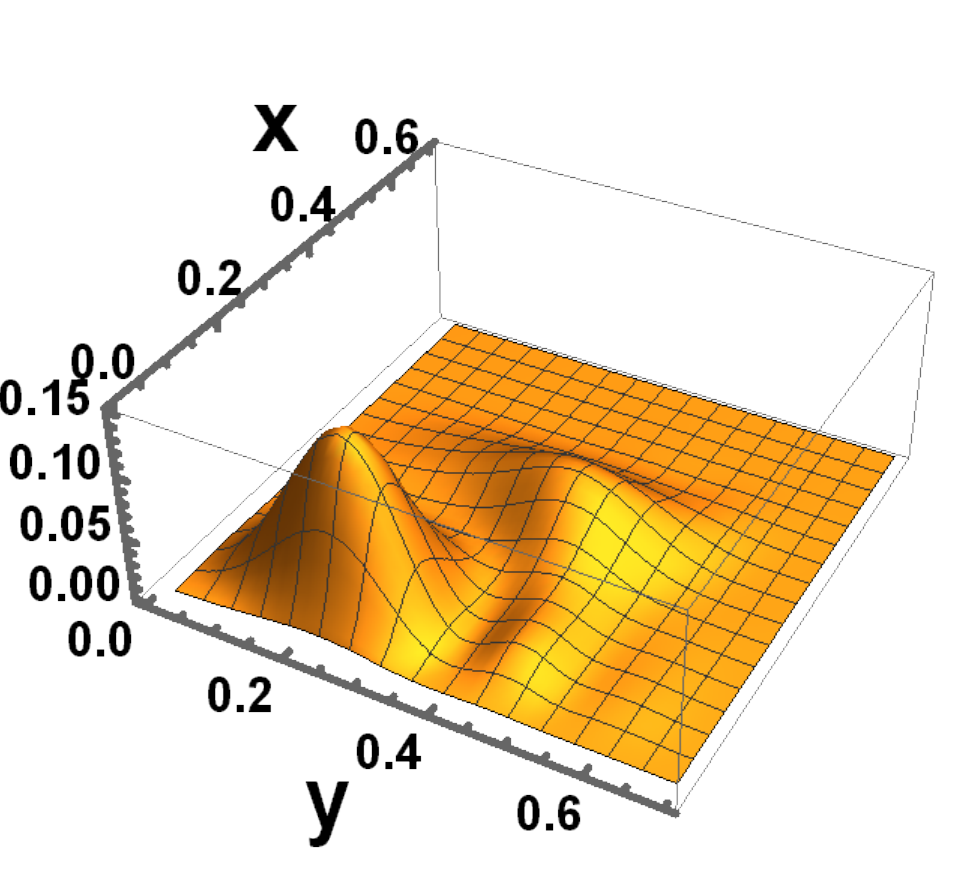} &
		\includegraphics[width=.15\textwidth]{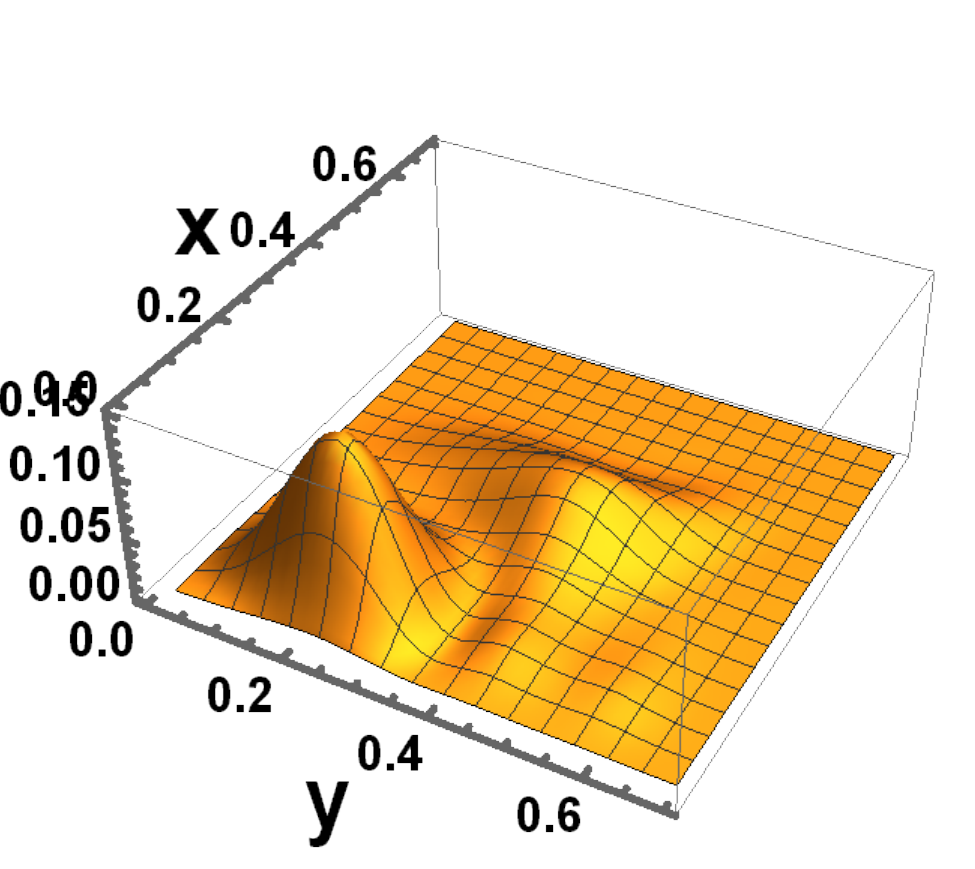} \\
	\end{tabular}
	\caption{(Color online) Evolution of the probability density $\int d\varphi|\psi(x,y,\varphi,t)|^2$ integrated over the angular variable $\varphi$. The system is prepared in the $(1,0)$ excited state. The initial interaction $g_{\textmd{in}}=-1$ is quenched to $g_{\textmd{f}}=1$.}\label{fwd_wf}
\end{figure}

Let us have a look at Fig.~\ref{fwd_wf}, which displays the evolution of the probability density, integrated over the angular variable $\varphi$, within the time window $t\in[16.4,18]$. At $t=16.4$, the probability density exhibits two humps. The positioning of these humps closely resembles that of the $(0,1)$ prequench excited state for $g=-1$ (cf. Fig.~\ref{wf}) with respect to the relative motion. At later times, a new hump emerges near the origin, causing the redistribution of height and width among the previous two humps. The position of this new hump, along with the noticeable gap (which we interpret as a nodal line) between it and the initial hump along the y-axis, serves as a clear indication of the emergence of the center-of-mass excited state. By $t=17.2$, the presence of three distinct humps becomes evident. Subsequently, the third hump continues to grow, while the initial hump significantly diminishes by $t=18$.

\subsection{Quench dynamics from the repulsive to the attractive interaction}

In this section, we examine the reverse scenario where the initial state corresponds to a repulsive interaction, $g_{\textmd{in}}=1$.

\begin{figure}[H]
	\centering
	\includegraphics[width=8.5cm,clip]{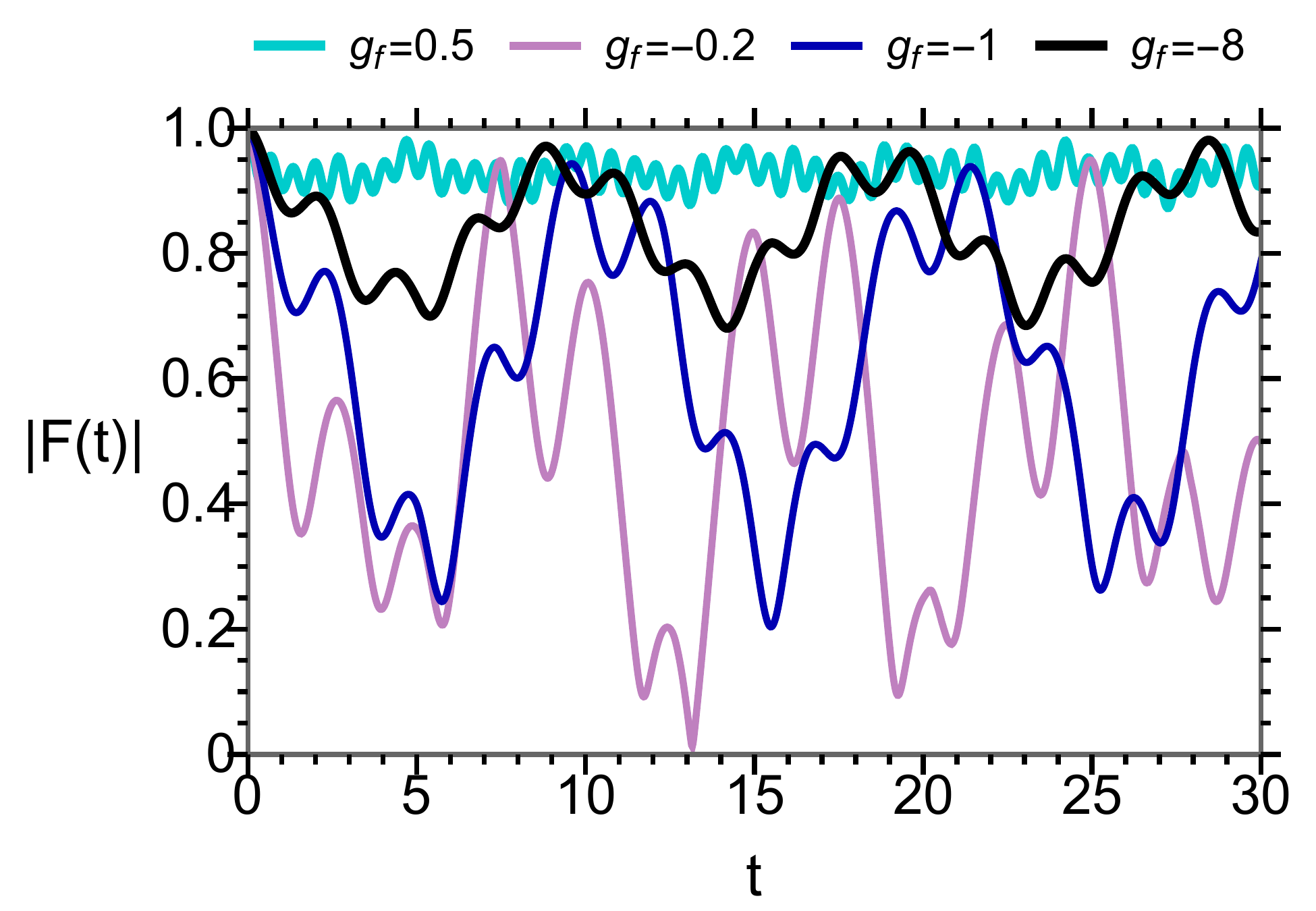}
	\caption{(Color online) Fidelity $F(t)$ for different values of the coupling constant $g_{\textmd{f}}$ in the case of the anisotropic harmonic trap. The initial coupling constant is $g_{\textmd{in}}=1$.}\label{bwd_2}
\end{figure}

For comparison purposes, we adjust the coupling strength $g_{\textmd{f}}$ to the same set of values as in the isotropic case. Fig.~\ref{bwd_2} displays the fidelity $|F(t)|$ for these values. Comparing it with Fig.~\ref{bwd_ho}, we observe significant distortion in the behavior of $|F(t)|$ in this anisotropic case. Nevertheless, we notice the involvement of more than one frequency, in contrast to the isotropic case. The oscillation amplitude and period become larger. The overall tendency, however, exhibits some similarity. It starts with a lower oscillation amplitude for $g_{\textmd{f}}=0.5$, followed by an increase, as observed for $g_{\textmd{f}}=-0.2$. Subsequently, it decreases again for $g_{\textmd{f}}=-1$ and $g_{\textmd{f}}=-8$.

\begin{figure}%[H]
	\centering
%	\textbf{$\alpha=0$}\par\medskip
	\includegraphics[width=8.5cm,clip]{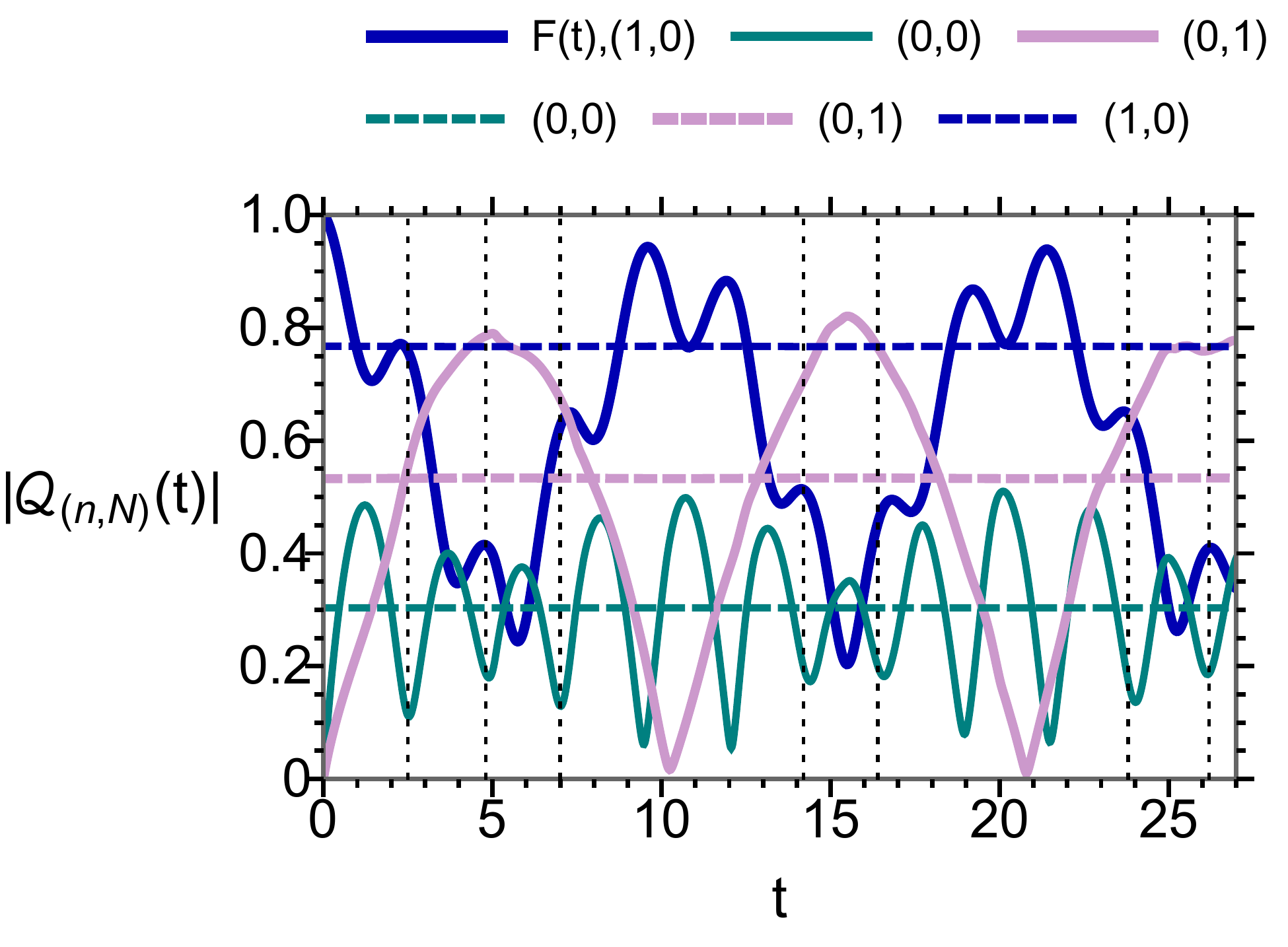}
	\caption{(Color online) Fidelity $F(t)$ and the overlap integrals $\mathcal{Q}$ between the time-evolving state $\psi(t)$ and different pre- (solid lines) and postquench (dashed lines) states in the case of the anisotropic harmonic trap. The indices $(n,N)$ refer to the states with the quantum numbers of the relative and center-of-mass motions. The system of the two atoms is prepared in the $(1,0)$ excited state with $g_{\textmd{in}}=1$ and quenched to $g_{\textmd{f}}=-1$. The dashed vertical lines indicate the emergence of the molecular excited state - $(0,1)$.}\label{bwd}
\end{figure}

We now examine the dynamics with a fixed interaction strength of $g_{\textmd{f}}=-1$ for $t>0$. Fig.~\ref{bwd} illustrates the fidelity $F(t)$ along with the overlaps between the evolving state and various prequench (solid curves) and postquench states (dashed lines). From the figure, two oscillating frequencies, a faster and a slower one, are noticeable in the fidelity $F(t)$. At the minimums of the oscillation period of $F(t)$, corresponding to the slower frequency, the overlap with the $(0,1)$ prequench excited state reaches its maximum value. It is this time window when the system occupies the molecular excited state. We mark these occurrences with vertical dashed lines. Additionally, the overlap with the $(0,1)$ postquench excited state also exhibits a high value, indicating the emergence of the molecular excited state. However, it is the overlap with the $(0,1)$ prequench excited state that enables the approximate detection of the time moments when such a state emerges.

To observe the emergence of the molecular excited state in terms of the evolution of the probability density, we focus on the time interval $t\in[3.2,6]$. At $t=3.2$, the probability density, integrated over the angular variable $\varphi$, closely resembles the integrated probability density of the $(1,0)$ relative excited state for $g=-1$ (cf. Fig.~\ref{wf}). Subsequently, one of the humps in the probability density begins to elongate towards the origin, resulting in a configuration at $t=4.8$ where it becomes approximately parallel to another hump, leading to a noticeable gap between them. This formation of new humps bears resemblance to the hump formation observed for the $(0,1)$ center-of-mass (molecular) excited state. By $t=6$, the initial configuration is nearly restored.

\begin{figure}[H]
	\centering
	\begin{tabular}{ccc}
		$t=3.2$ & $t=3.6$ & $t=4$ \\
		\includegraphics[width=.15\textwidth]{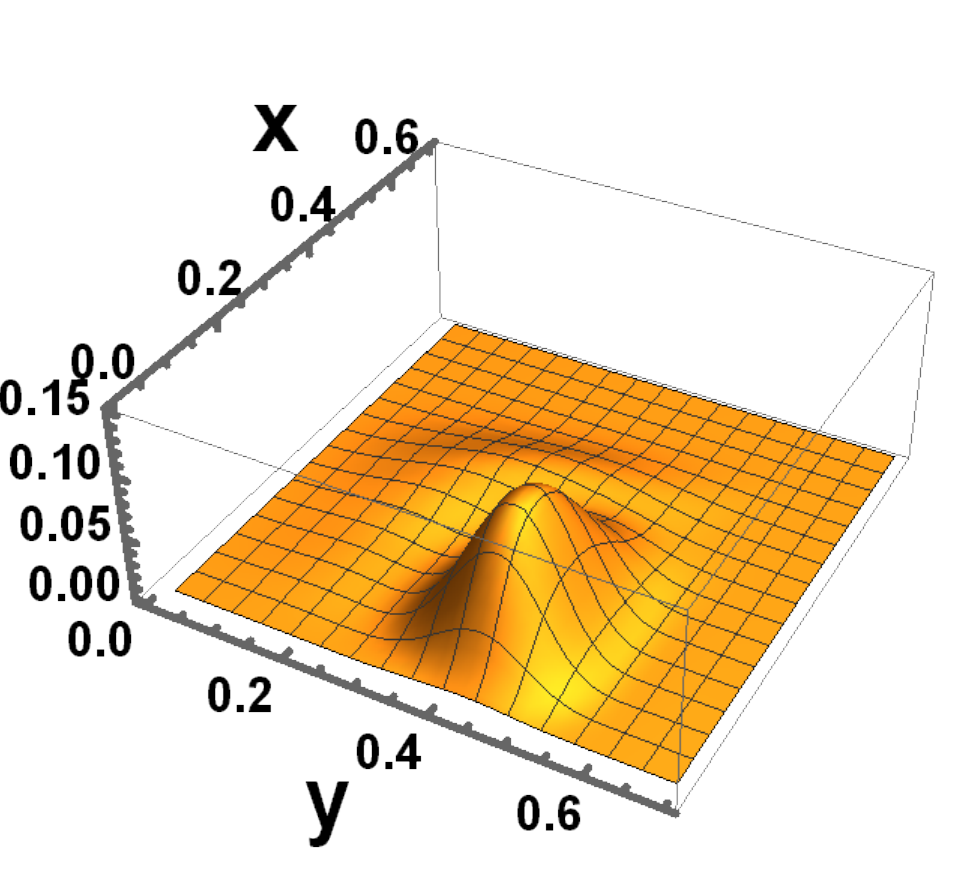} &
		\includegraphics[width=.15\textwidth]{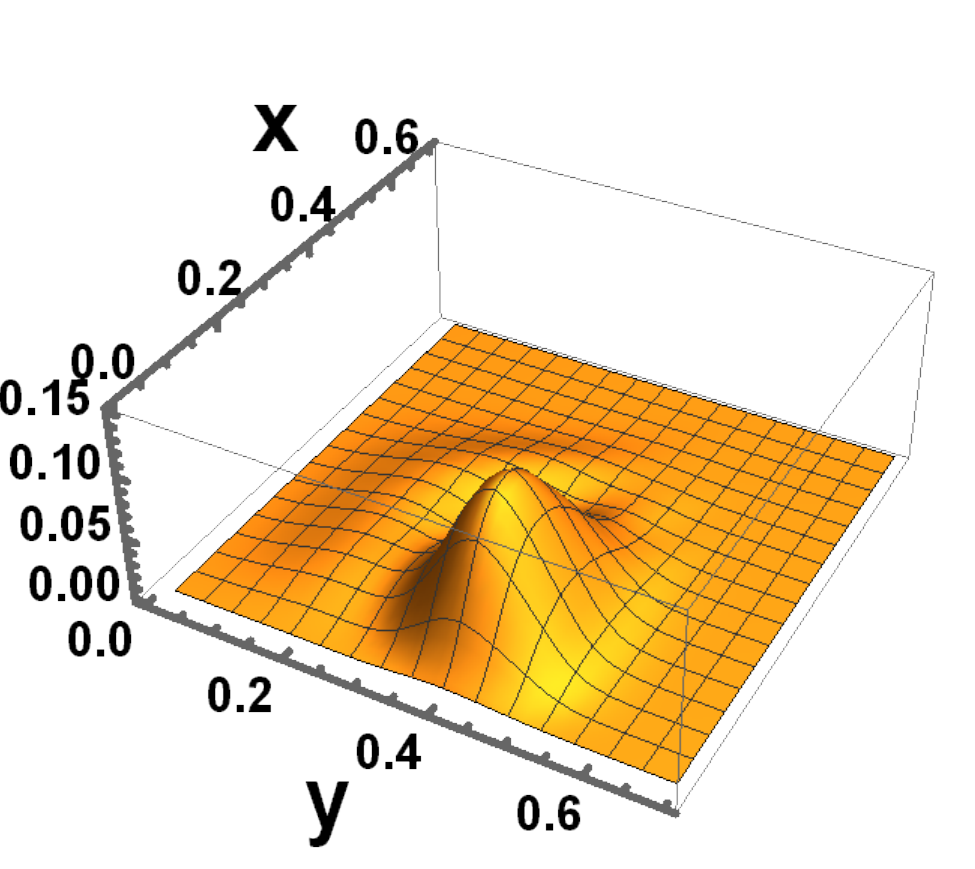} &
		\includegraphics[width=.15\textwidth]{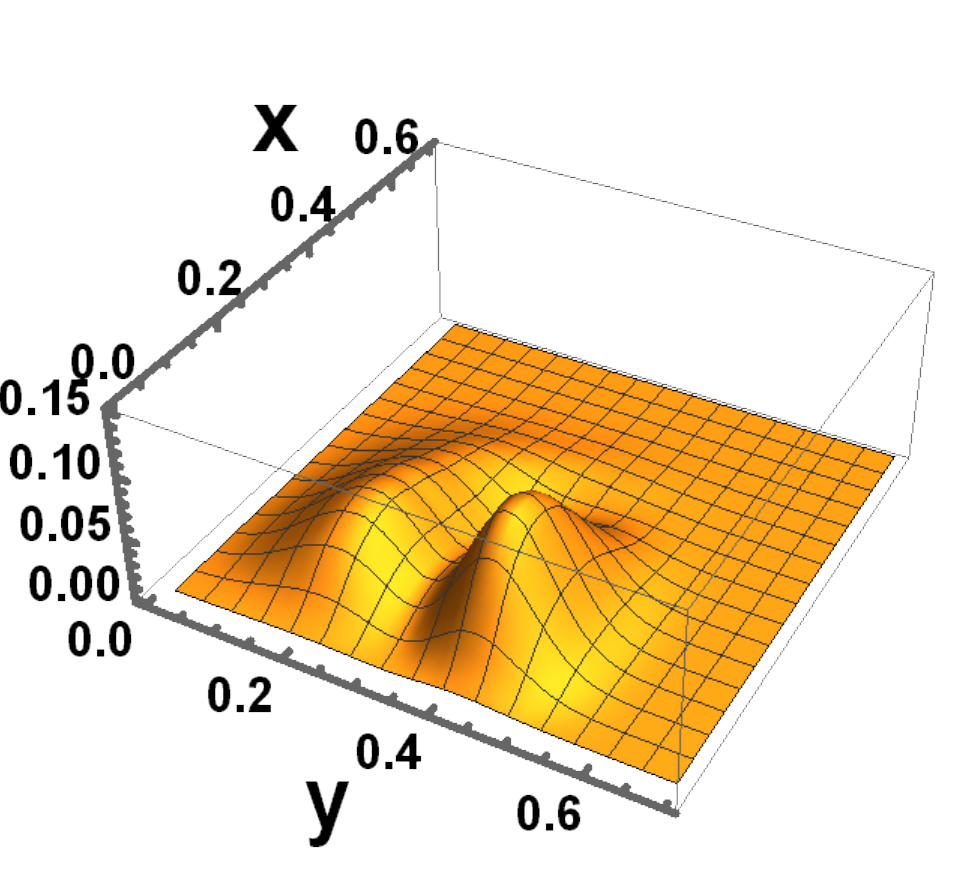} \\
		$t=4.8$ & $t=5.6$ & $t=6$\\
		\includegraphics[width=.15\textwidth]{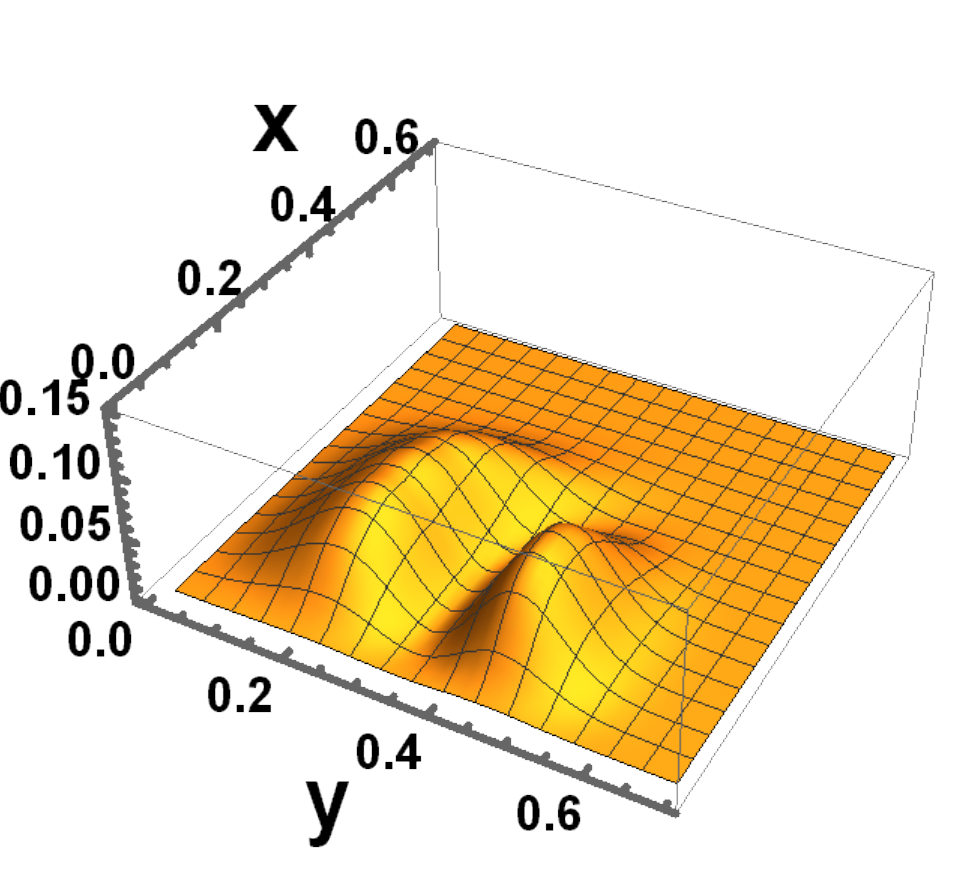} &
		\includegraphics[width=.15\textwidth]{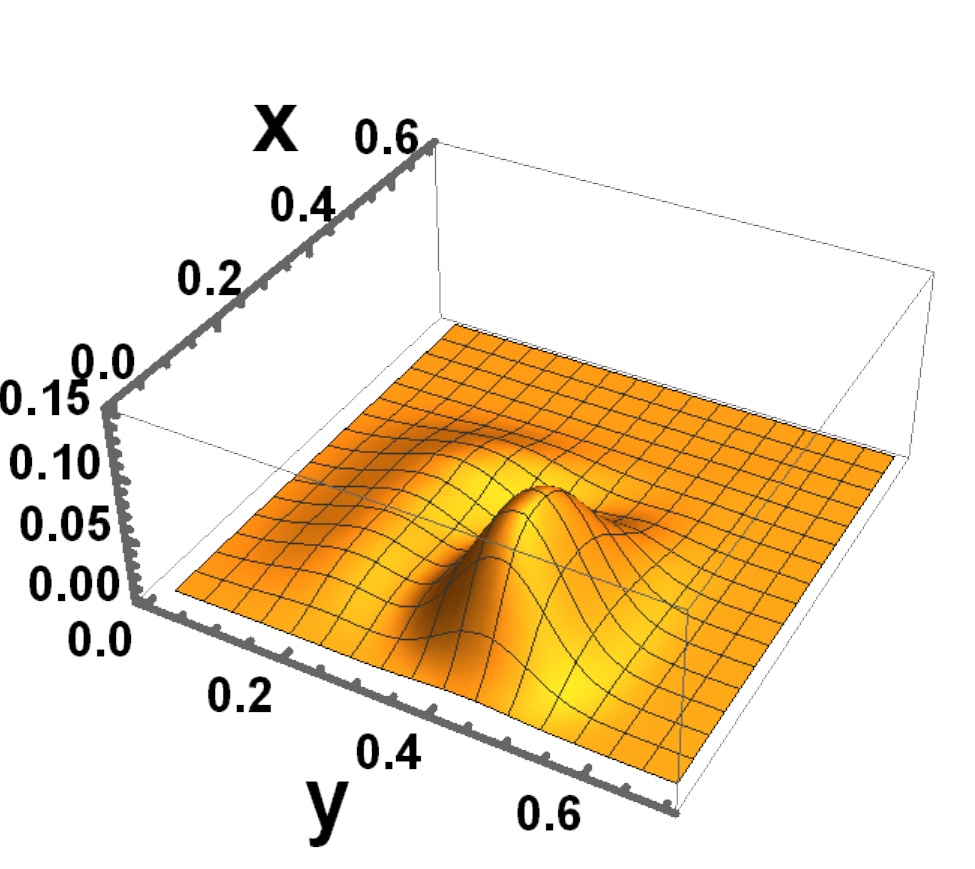} &
		\includegraphics[width=.15\textwidth]{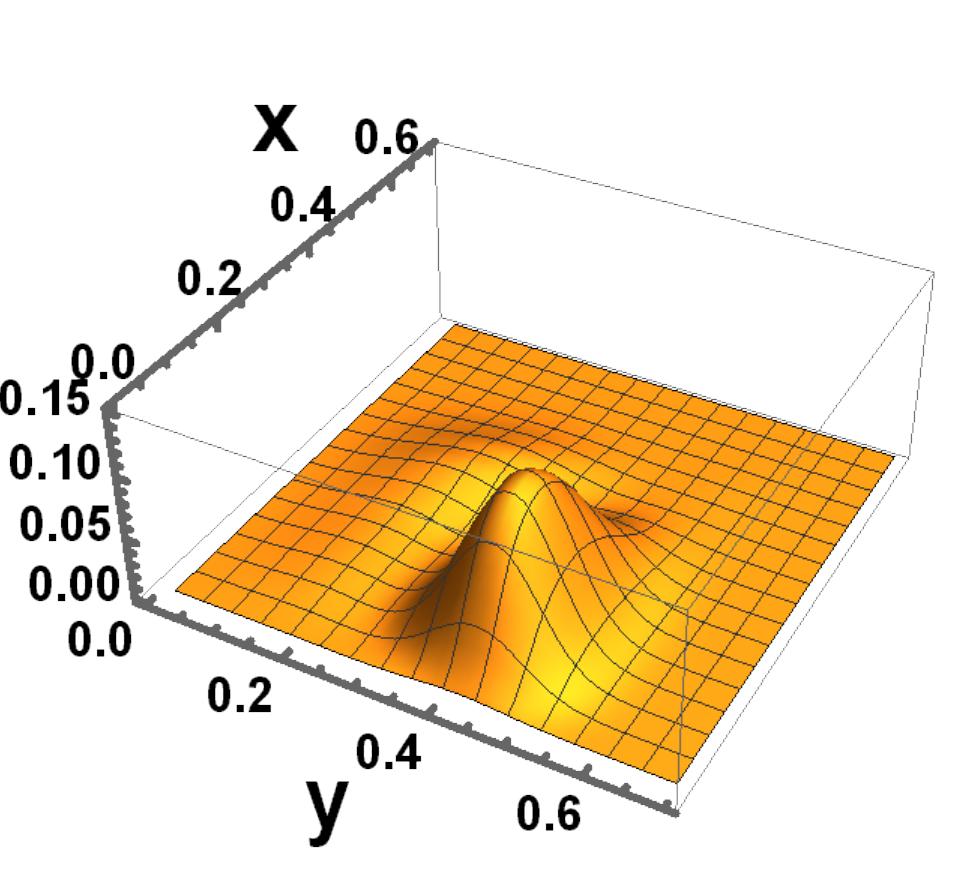} \\
	\end{tabular}
	\caption{(Color online) Evolution of the probability density $\int d\varphi|\psi(x,y,\varphi,t)|^2$ integrated over the angular variable $\varphi$. The system is prepared in the $(1,0)$ excited state. The initial interaction $g_{\textmd{in}}=1$ is quenched to $g_{\textmd{f}}=-1$.}\label{bwd_wf}
\end{figure}

\section{Summary}\label{summary}

An analysis of the quench dynamics involving two interacting atoms confined in an anisotropic trap was conducted. The interaction potential we considered was a finite-range interaction with a Gaussian shape. Initially, we focused on the isotropic harmonic trap and adjusted our finite-range interaction to match the energy levels obtained analytically in \cite{bougas} for the zero-range interaction. We computed the energy spectrum for the lowest levels in both the isotropic and anisotropic traps. In the case of the isotropic trap, the quench dynamics almost coincide with those observed for the zero-range interaction. This similarity allowed for a consistent representation of our results in terms of the coupling constant of the zero-range interaction. In the anisotropic case, the two atoms had different masses and were subjected to the harmonic traps with different frequencies. This case resulted in the emergence of molecular excited states during the quench dynamics when transitioning from attractive to repulsive interactions, and vice versa. Notably, when starting from the attractive interaction, a larger number of such occurrences were observed compared to the case where the atoms were initially in the repulsive regime. When the coupling constant is altered, the fidelity in the anisotropic case exhibits larger oscillation amplitudes and periods, resulting in a distorted overall behavior compared to the case of the isotropic trap.

As potential future prospects, utilizing the developed numerical method for the two-dimensional geometry, it would be interesting to investigate quench dynamics for various trapping potentials, such as a double-well potential \cite{koscik}, for two-component mass‑imbalance systems \cite{nandy}, for polar and paramagnetic molecules \cite{dawid} and for various types of interactions \cite{koscik2,koscik3}. Additionally, exploring dynamical excitation processes of two-dimensional mixtures \cite{bougas2,bougas3} and properties of systems in a dimensional crossover \cite{bougas4} could provide valuable insights.

\section*{Acknowledgment}

We would like to express our gratitude to \mbox{Prof. V. S. Melezhik} for his insightful explanation on solving the time-dependent Schrödinger equation, which forms the basis of the content discussed in Section \ref{model}.

We also thank G. Bougas for providing us with the data for the zero-range interaction and discussing some related aspects of it.

This research has been funded by the Science Committee of the Ministry of Education and Science of the Republic of Kazakhstan (Grant No. AP13067639).

This preprint has not undergone peer review (when applicable) or any post-submission improvements or corrections. The Version of Record of this article is published in The European Physical Journal Plus, and is available online at https://doi.org/10.1140/epjp/s13360-024-04864-2

\section*{Data availability statement}

All data is available upon reasonable request.


\begin{thebibliography}{99}

\bibitem{hartke} T. Hartke, B. Oreg, N. Jia, and M. Zwierlein, \textit{Quantum Register of Fermion Pairs}, Nature \textbf{601}, 537–541 (2022).

\bibitem{wu} Y.-K. Wu and L.-M. Duan, \textit{A Two-Dimensional Architecture for Fast Large-
	Scale Trapped-Ion Quantum Computing}, Chinese Phys. Lett. \textbf{37}, 070302 (2020).

\bibitem{bruzewicz} C. D. Bruzewicz, R. McConnell, J. Chiaverini, and J. M. Sage. \textit{Scalable loading of a two-dimensional trapped-ion array}, Nat Commun \textbf{7}, 13005 (2016).

\bibitem{busch} T. Busch, B. Englert, K. Rzazewski, and M. Wilkens, \textit{Two Cold Atoms in a Harmonic Trap}, Found. Phys. \textbf{28}, 549 (1998).

\bibitem{idziaszek} Z. Idziaszek, and T. Calarco, \textit{Analytical solutions for the dynamics of two trapped interacting ultracold atoms}, Phys. Rev. A \textbf{74}, 022712 (2006).

\bibitem{chen} Y. Chen, D.-W. Xiao, R. Zhang, and P. Zhang, \textit{Analytical solution for the spectrum of two ultracold atoms in a completely anisotropic confinement}, Phys. Rev. A \textbf{101}, 053624 (2020).

\bibitem{olshanii} M. Olshanii, \textit{Atomic Scattering in the Presence of an External Confinement and a Gas of Impenetrable Bosons}, Phys. Rev. Lett., \textbf{81}, 938 (1998).

\bibitem{melezhik2007} V. S. Melezhik, J. I. Kim, and P. Schmelcher, \textit{Wave-packet dynamical analysis of ultracold scattering in cylindrical waveguides}, Phys. Rev. A \textbf{76}, 053611 (2007).

\bibitem{melezhik2009} V. S. Melezhik, and P. Schmelcher, \textit{Quantum dynamics of resonant molecule formation in waveguides}, New J. Phys. \textbf{11}, 073031 (2009).

\bibitem{haller} E. Haller, M. J. Mark, R. Hart, J. G. Danzl, L. Reichsollner, V. Melezhik, P. Schmelcher, and H.-C. N\"agerl, \textit{Confinement-induced resonances in low-dimensional quantum systems}, Phys. Rev. Lett., \textbf{104}, 153203, (2010).

\bibitem{sala} S. Sala and A. Saenz, \textit{Theory of inelastic confinement-induced resonances due to the coupling of center-of-mass and relative motion}, Phys. Rev. A \textbf{94}, 022713 (2016).

\bibitem{peng} S.-G. Peng, H. Hu, X.-J. Liu, and P.D. Drummond, \textit{Confinement-Induced Resonances in Anharmonic Waveguides}, Phys. Rev. A \textbf{84}, 043619 (2011).

\bibitem{sala2} S. Sala, G. Zürn, T. Lompe, A. N. Wenz, S. Murmann, F. Serwane, S. Jochim, and A. Saenz, \textit{Coherent Molecule Formation in Anharmonic Potentials Near Confinement-Induced Resonances}, Phys. Rev. Lett. \textbf{110}, 203202 (2013).

\bibitem{gharashi} S. E. Gharashi and D. Blume, \textit{Tunneling dynamics of two interacting one-dimensional particles}, Phys. Rev. A. \textbf{92}, 033629 (2015).

\bibitem{dobrzyniecki} J. Dobrzyniecki and T. Sowi\'{n}ski, \textit{Two Rydberg-dressed atoms escaping from an open well}, Phys. Rev. A. \textbf{103}, 013304 (2021)

\bibitem{ishmukh} I. S. Ishmukhamedov and V. S. Melezhik, \textit{Tunneling of two bosonic atoms from a one-dimensional anharmonic trap}, Phys. Rev. A. \textbf{95}, 062701 (2017).

\bibitem{ishmukh2} I. S. Ishmukhamedov and A. S. Ishmukhamedov, \textit{Tunneling of two interacting atoms from excited states},  Physica E \textbf{109}, 24 (2019).

\bibitem{ishmukh3} I. S. Ishmukhamedov, \textit{Quench dynamics of two interacting atoms in a one-dimensional anharmonic trap},  Physica E \textbf{142}, 115228 (2022).

\bibitem{serwane} F. Serwane, G. Z\"{u}rn, T. Lompe, T. B. Ottenstein, A. N. Wenz, and S. Jochim, \textit{Deterministic preparation of a tunable few-fermion system}, Science \textbf{332}, 336 (2011).

\bibitem{zuern} G. Z\"{u}rn, F. Serwane, T. Lompe, A. N. Wenz, M. G. Ries, J. E. Bohn, and S. Jochim, \textit{Fermionization of Two Distinguishable Fermions}, Phys. Rev. Lett. \textbf{108}, 075303 (2012).

\bibitem{bayha} L. Bayha, M. Holten, R. Klemt, K. Subramanian, J. Bjerlin, S. M. Reimann, G. M. Bruun, P. M. Preiss, and S. Jochim \textit{Observing the emergence of a quantum phase transition shell by shell}, Nature \textbf{587}, 583–587 (2020).

\bibitem{holten} M. Holten, L. Bayha, K. Subramanian, S. Brandstetter, C. Heintze, P. Lunt, P. M. Preiss, and S. Jochim \textit{Observation of Cooper pairs in a mesoscopic two-dimensional Fermi gas}, Nature \textbf{606}, 287–291 (2022).

\bibitem{langen} T Langen, R Geiger, J Schmiedmayer, \textit{Ultracold atoms out of equilibrium}, Annu. Rev. Condens. Matter Phys. \textbf{6} (1), 201-217 (2015).

\bibitem{bougas} G. Bougas, S. I. Mistakidis, and P. Schmelcher \textit{Analytical treatment of the interaction quench dynamics of two bosons in a two-dimensional harmonic trap}, Phys. Rev. A \textbf{100}, 053602 (2019)

\bibitem{ishmukh4} I. S. Ishmukhamedov, D. S. Valiolda, and S. A. Zhaugasheva, \textit{Description of ultracold atoms in a one-dimensional geometry of a harmonic trap with a realistic interaction}, Phys. Part. Nuclei Lett. \textbf{11}: 238 (2014).

\bibitem{ishmukh5} I. S. Ishmukhamedov, D. T. Aznabayev, and S. A. Zhaugasheva, \textit{Two-body atomic system in a one-dimensional anharmonic trap: The energy spectrum},  Phys. Part. Nucl. Lett. \textbf{12}, 680 (2015).

\bibitem{doganov} R. A. Doganov, S. Klaiman, O. E. Alon, A. I. Streltsov, and L. S. Cederbaum, \textit{Two trapped particles interacting by a finite-range two-body potential in two spatial dimensions}, Phys. Rev. A \textbf{87}, 033631 (2013).

\bibitem{bock} H. Bock, I. Lesanovsky, and P. Schmelcher, \textit{Neutral two-body systems in inhomogeneous magnetic fields: the quadrupole configuration}, J. Phys. B: At. Mol. Opt. Phys. \textbf{38}, 893–906 (2005)

\bibitem{borisov} A. G. Borisov, \textit{Solution of the radial Schrödinger equation in cylindrical and spherical coordinates by mapped Fourier transform algorithms}, J. Chem. Phys. \textbf{114}, 7770 (2001).

\bibitem{lemoine} D. Lemoine, \textit{Optimized grid representations in curvilinear coordinates: the mapped sine Fourier method}, Chem. Phys. Lett. \textbf{320}, 492 (2000)

\bibitem{melezhik_pc} V. S. Melezhik (private communication)

\bibitem{koval} E. A. Koval, O. A. Koval, and V. S. Melezhik, \textit{Anisotropic quantum scattering in two dimensions}, Phys. Rev. A \textbf{89}, 052710 (2014).

\bibitem{melezhik} V. S. Melezhik, \textit{New method for solving multidimensional scattering problem}, J. Comput. Phys. \textbf{92}, 67 (1991).

\bibitem{shadmehri} S. Shadmehri, and V. S. Melezhik, \textit{A hydrogen atom in strong elliptically polarized laser fields within discrete variable representation}, Laser Phys. \textbf{33}, 026001 (2023)

\bibitem{ishmukh6} I. Ishmukhamedov, A. Ishmukhamedov, and V. Melezhik, \textit{Numerical Solution of the Time Dependent 3D Schr\"odinger Equation Describing Tunneling of Atoms from Anharmonic Traps},   	
EPJ Web Conf. \textbf{173}, 03011 (2018)

\bibitem{bougas_data} The data for the zero-range interaction were kindly provided to us in tabular form by G. Bougas.

\bibitem{bougas_pc} G. Bougas (private communication)

\bibitem{koscik} P. Ko\'{s}cik, \textit{On the Exponential Decay of Strongly Interacting Cold Atoms from a Double-Well Potential}, Few-Body Syst. \textbf{64}, 11 (2023)

\bibitem{nandy} D. K. Nandy, and T. Sowi\'{n}ski, \textit{Sudden quench of harmonically trapped mass‑imbalanced fermions}, Sci. Rep. \textbf{12}, 19710 (2022).

\bibitem{dawid} A. Dawid, and M. Tomza, \textit{Magnetic properties and quench dynamics of two interacting ultracold molecules in a trap}, Phys. Chem. Chem. Phys. \textbf{22}, 28140 (2020).

\bibitem{koscik2} P. Ko\'{s}cik, and T. Sowi\'{n}ski, \textit{Exactly solvable model of two interacting Rydberg-dressed atoms confined in a two-dimensional harmonic trap}, Sci. Rep. \textbf{9}, 12018 (2019).

\bibitem{koscik3} P. Ko\'{s}cik, and T. Sowi\'{n}ski, \textit{Universality of Internal Correlations of Strongly Interacting -Wave Fermions in One-Dimensional Geometry}, Phys. Rev. Lett. \textbf{130}, 253401 (2023).

\bibitem{bougas2} G. Bougas, S. I. Mistakidis, P. Giannakeas, and P. Schmelcher, \textit{Dynamical excitation processes and correlations of three-body two-dimensional mixtures}, Phys. Rev. A \textbf{106}, 043323 (2022).

\bibitem{bougas3} G. Bougas, S. I. Mistakidis, P. Giannakeas, and P. Schmelcher, \textit{Few-body correlations in two-dimensional Bose and Fermi ultracold mixtures}, New J. Phys. \textbf{23}, 093022 (2021).

\bibitem{bougas4} G. Bougas, S. I. Mistakidis, G. M. Alshalan, and P. Schmelcher, \textit{Stationary and dynamical properties of two harmonically trapped bosons in the crossover from two dimensions to one}, Phys. Rev. A \textbf{102}, 013314 (2020).
  
\end{thebibliography}
\end{document}